\documentclass[usenatbib]{mnras}
\usepackage{graphicx}
\usepackage{amsmath}
\usepackage{amssymb}
\usepackage{color}
\usepackage{url}
\usepackage{CJKutf8}
\usepackage{array}

\usepackage{threeparttable}

\newcommand\lsim{\mathrel{\rlap{\lower4pt\hbox{\hskip1pt$\sim$}}
        \raise1pt\hbox{$<$}}}
\newcommand\gsim{\mathrel{\rlap{\lower4pt\hbox{\hskip1pt$\sim$}}
        \raise1pt\hbox{$>$}}}
\newcommand{\lya}{\ifmmode\mathrm{Ly}\alpha\else{}Ly$\alpha$\fi}
\newcommand{\lyb}{\ifmmode\mathrm{Ly}\beta\else{}Ly$\beta$\fi}
\newcommand{\igm}{\ifmmode\mathrm{IGM}\else{}IGM\fi}
\newcommand{\lae}{\ifmmode\mathrm{LAE}\else{}LAE\fi}
\newcommand{\h}{\ifmmode\mathrm{H}\else{}H\fi}
\newcommand{\hi}{\ifmmode\mathrm{H\,{\scriptscriptstyle I}}\else{}H\,{\scriptsize I}\fi}
\newcommand{\hii}{\ifmmode\mathrm{H\,{\scriptscriptstyle II}}\else{}H\,{\scriptsize II}\fi}
\newcommand{\cmb}{\ifmmode\mathrm{CMB}\else{}CMB\fi}
\newcommand{\qso}{\ifmmode\mathrm{QSO}\else{}QSO\fi}
\newcommand{\eor}{\ifmmode\mathrm{EoR}\else{}EoR\fi}
\newcommand{\heii}{\ifmmode\mathrm{He\,{\scriptscriptstyle II}}\else{}He\,{\scriptsize II}\fi}
\newcommand{\heiii}{\ifmmode\mathrm{He\,{\scriptscriptstyle III}}\else{}He\,{\scriptsize III}\fi}
\newcommand{\ciii}{\ifmmode\mathrm{C\,{\scriptscriptstyle III]}}\else{}C\,{\scriptsize III]}\fi}
\newcommand{\oiii}{\ifmmode\mathrm{O\,{\scriptscriptstyle III}}\else{}O\,{\scriptsize III}\fi}
\newcommand{\aliii}{\ifmmode\mathrm{Al\,{\scriptscriptstyle III}}\else{}Al\,{\scriptsize III}\fi}
\newcommand{\mgii}{\ifmmode\mathrm{Mg\,{\scriptscriptstyle II}}\else{}Mg\,{\scriptsize II}\fi}
\newcommand{\fe}{\ifmmode\mathrm{Fe}\else{}Fe\fi}
\newcommand{\nv}{\ifmmode\mathrm{N\,{\scriptscriptstyle V}}\else{}N\,{\scriptsize V}\fi}
\newcommand{\niv}{\ifmmode\mathrm{N\,{\scriptscriptstyle IV]}}\else{}N\,{\scriptsize IV]}\fi}
\newcommand{\cii}{\ifmmode\mathrm{C\,{\scriptscriptstyle II}}\else{}C\,{\scriptsize II}\fi}
\newcommand{\civ}{\ifmmode\mathrm{C\,{\scriptscriptstyle IV}}\else{}C\,{\scriptsize IV}\fi}
\newcommand{\siv}{\ifmmode\mathrm{Si\,{\scriptscriptstyle IV}}\else{}Si\,{\scriptsize IV}\fi}
\newcommand{\siii}{\ifmmode\mathrm{Si\,{\scriptscriptstyle II}}\else{}Si\,{\scriptsize II}\fi}
\newcommand{\siiii}{\ifmmode\mathrm{Si\,{\scriptscriptstyle III]}}\else{}Si\,{\scriptsize III]}\fi}
\newcommand{\ovi}{\ifmmode\mathrm{O\,{\scriptscriptstyle VI}}\else{}O\,{\scriptsize VI}\fi}
\newcommand{\sioiv}{\ifmmode\mathrm{Si\,{\scriptscriptstyle IV}\,\plus O\,{\scriptscriptstyle IV]}}\else{}Si\,{\scriptsize IV}\,+O\,{\scriptsize IV]}\fi}

\newcommand{\cmmc}{\textsc{\small 21CMMC}}
\newcommand{\cmfst}{\textsc{\small 21CMFAST}}

 \makeatletter
 \newcommand*{\@rowstyle}{}
\newcommand*{\rowstyle}[1]{
 \gdef\@rowstyle{#1}%
 \@rowstyle\ignorespaces%
}
\newcolumntype{=}{
>{\gdef\@rowstyle{}}%
}
\newcolumntype{+}{
>{\@rowstyle}%
}

 \newcolumntype{C}[1]{>{\centering\arraybackslash}p{#1}}
 \makeatother

\pdfoutput=1
\title[Wavelet Scattering Transform applied to the EoR]{Exploring the cosmic 21-cm signal from the Epoch of Reionisation using the Wavelet Scattering Transform}
\author[B. Greig et al.] {Bradley~Greig$^{1,2}$\thanks{E-mail:~greigb@unimelb.edu.au}, Yuan-Sen Ting (丁源森)$^{2,3,4}$ \& Alexander A. Kaurov$^{5}$ \\
$^1$School of Physics, University of Melbourne, Parkville, VIC 3010, Australia \\
$^2$ARC Centre of Excellence for All-Sky Astrophysics in 3 Dimensions (ASTRO 3D) \\
$^3$Research School of Astronomy \& Astrophysics, Australian National University, Cotter Road, Weston Creek, ACT 2611, Canberra, Australia \\
$^4$Research School of Computer Science, Australian National University, Acton ACT 2601, Australia \\
$^5$Institute for Advanced Study, 1 Einstein Drive, Princeton, NJ 08540, USA \\
}

\begin{document}
\label{firstpage}
\pagerange{\pageref{firstpage}--\pageref{lastpage}}
\begin{CJK}{UTF8}{gkai} 
\maketitle
\end{CJK}

\begin{abstract}
\noindent

Detecting the cosmic 21-cm signal during the Epoch of Reionisation and Cosmic Dawn will reveal insights into the properties of the first galaxies and advance cosmological parameter estimation. Until recently, the primary focus for astrophysical parameter inference from the 21-cm signal centred on the power spectrum (PS). However, the cosmic 21-cm signal is highly non-Gaussian rendering the PS sub-optimal for characterising the cosmic signal. In this work, we introduce a new technique to analyse the non-Gaussian information in images of the 21-cm signal called the Wavelet Scattering Transform (WST). This approach closely mirrors that of convolutional neural networks with the added advantage of not requiring tuning or training of a neural network. Instead, it compresses the 2D spatial information into a set of coefficients making it easier to interpret while also providing a robust statistical description of the non-Gaussian information contained in the cosmic 21-cm signal. First, we explore the application of the WST to mock 21-cm images to gain valuable physical insights by comparing to the known behaviour from the 21-cm PS. Then we quantitatively explore the WST applied to the 21-cm signal by extracting astrophysical parameter constraints using Fisher Matrices from a realistic 1000 hr mock observation with the Square Kilometre Array. We find that: (i) the WST applied only to 2D images can outperform the 3D spherically averaged 21-cm PS, (ii) the excision of foreground contaminated modes can degrade the constraining power by a factor of $\sim1.5-2$ with the WST and (iii) higher cadences between the 21-cm images can further improve the constraining power.
\end{abstract} 


\section{Introduction}

Following recombination, the early Universe is rendered opaque to most forms of radiation owing to the pervasive neutral hydrogen fog. It remains in this state until the ignition of star-formation in the first primordial galaxies begins to locally ionise the neutral hydrogen, referred to as the Cosmic Dawn (CD). Over time, as these galaxies become larger, more abundant and clustered the cumulative ionising radiation from these sources eventually renders the intergalactic medium (IGM) almost completely ionised, referred to as the Epoch of Reionisation (EoR). 

The most promising tool at our disposal for probing the EoR and CD is the 21-cm hyperfine spin-flip transition of neutral hydrogen. This signal, measured relative to a background radiation source, such as the Cosmic Microwave Background \citep[see e.g.][]{Gnedin:1997p4494,Madau:1997p4479,Shaver:1999p4549,Tozzi:2000p4510,Gnedin:2004p4481,Furlanetto:2006p209,Morales:2010p1274,Pritchard:2012p2958}, can be seen in either emission or absorption depending on both the thermal and ionisation states of the IGM. By measuring the time-evolution (frequency dependence) of the 21-cm radiation along with its spatial distribution in the IGM, we are able to build up a three dimensional movie of the IGM in the early Universe. With this, we can infer the typical properties of the galaxies responsible for driving the EoR and CD.

Detecting this three dimensional 21-cm signal requires large-scale radio interferometer experiments, which are specifically designed to be sensitive to the spatial fluctuations. These naturally operate in the Fourier domain, resulting in the power spectrum (PS) being the preferred statistic to characterise the 21-cm signal. However, the 21-cm signal is non-Gaussian owing to the complex nature of the ionisation morphology. Thus, the PS is sub-optimal as it completely disregards the important phase information in the signal (i.e. the locations of the ionised regions). For the first generation of interferometer experiments, such as the Low-Frequency Array (LOFAR; \citealt{vanHaarlem:2013p200}), the Murchison Wide Field Array (MWA; \citealt{Tingay:2013p2997,Wayth:2018}), the Precision Array for Probing the Epoch of Reionisation (PAPER; \citealt{Parsons:2010p3000}), the Owens Valley Radio Observatory Long Wavelength Array (OVRO-LWA; \citealt{Eastwood:2019}) and the upgraded Giant Metrewave Radio Telescope (uGMRT; \citealt{Gupta:2017}) focusing on the PS is not a concern as these are likely sensitive to only a low signal-to-noise detection of the 21-cm signal. However, next generation experiments such as the Hydrogen Epoch of Reionization Array (HERA; \citealt{DeBoer:2017p6740}), NenuFAR (New extension in Nan\c{c}ay Upgrading loFAR; \citealt{Zarka:2012}) and the Square Kilometre Array (SKA; \citealt{Mellema:2013p2975,Koopmans:2015}) should yield sufficiently high sensitivity to be able to measure the 21-cm signal beyond the PS. Further, the SKA has been specifically designed to be able to provide the first three-dimensional tomographic images of the 21-cm signal completely bypassing the need to analyse the signal in the Fourier domain.

Recognising this, in recent years several alternative non-Gaussian probes of the 21-cm signal have been explored. The natural extension to the PS in the presence of non-Gaussian information is to consider the Bispectrum, the Fourier transform of the three-point correlation function. The sensitivity of the Bispectrum to the 21-cm signal has been extensively studied in the literature \citep[e.g.][]{Yoshiura:2015,Shimabukuro:2016,Majumdar:2018,Watkinson:2019,Hutter:2020,Majumdar:2020,Kamran:2021} along with closely related approaches such as the triangle correlation function \citep[focussing on the phase information only;][]{Gorce:2019} and the position dependent power spectrum \citep{Giri:2019b}. In accessing this non-Gaussian information, the Bispectrum has been shown to improve over that from the PS for astrophysical parameter recovery from a mock 21-cm signal \citep{Shimabukuro:2017,Tiwari:2021,Watkinson:2021}.

Beyond the Fourier domain, numerous alternative approaches to analyse images of the 21-cm signal have been explored. These include the one-point statistics of the brightness temperature \citep{Watkinson:2014,Shimabukuro:2015,Kubota:2016,Banet:2021,Gorce:2021}, the morphological and/or topological features of the 21-cm signal \citep[e.g.][]{Yoshiura:2017,Bag:2019,Chen:2019,Elbers:2019,Kapahtia:2019,Gazagnes:2021,Giri:2021,Kapahtia:2021} and the distribution of the sizes of ionised regions \citep{Kakiichi:2017,Giri:2018a,Giri:2018b,Giri:2019a,Bianco:2021}. Alternatively, convolutional neural networks (CNNs) have been considered which are trained to be able to extract 2D or 3D features from images of the cosmic 21-cm signal and used to extract astrophysical information from mock observations \citep[e.g.][]{Gillet:2019,Hassan:2019,LaPlante:2019,Hassan:2020,Kwon:2020,Mangena:2020,Prelogovic:2021}.

In this work, we introduce a new technique sensitive to the non-Gaussian information contained in images called the Wavelet Scattering Transform (WST). Specifically for this work, we focus on the WST only in two dimensions. First introduced by \citet{Mallat:2012}, the WST was used for signal processing of high-dimensional datasets and has since been applied to several different areas of astronomy, for example, for the interstellar medium \citep{Allys:2019,Blancard:2020,Saydjari:2021}, weak lensing \citep{Cheng:2020,Cheng:2021} and large-scale structure \citep{Allys:2020,Valogiannis:2021}. The WST applies a family of wavelet filters to an input image, extracting spatial features on different physical scales. Wavelets preserve the locality of the signal both spatially and in frequency (unlike the Fourier transform), ensuring the non-Gaussian information of the 21-cm signal is retained. By performing iterative convolutions of the input image by these filters, we measure the clustering of the spatial features, characterising the non-Gaussian information. Importantly for this work, in using pre-defined values for the filters, the WST is capable of characterising the global statistical properties of the signal by compressing the non-Gaussian information into a manageable set of scattering coefficients for performing robust statistical analyses.

While the bispectrum is also capable of measuring the non-Gaussian information, it suffers from the successive multiplication of the input information. As such, the variance of the statistic increases rapidly and is sensitive to outliers in the data. The WST is capable of circumventing this by taking the modulus after each successive filtering ensuring stability in the statistic (i.e. no divergence in the noise of the statistic). As a result, since the scattering coefficients are binning different Fourier modes, the variance of the scattering coefficients should approximately be Gaussian. The WST also shares features with CNNs, where wavelet filters are used to analyse the signal. In CNNs, the spatial scales and rotations of the wavelets are learnt by the training of the network, with the activation functions and pooling consistent with the modulus and reduction to scattering coefficients. On the other hand since the WST uses a pre-defined set of wavelets we do not require any complex neural network training. Further, since they are pre-defined, the information is easier to interpret. For more in-depth discussions between these approaches we refer the reader to \citet{Cheng:2021b}.  

First, to gain valuable insight into the WST and its application to the 21-cm signal, we apply the WST to a series of 21-cm simulations with different characteristics. Next, we progressively add in instrumental effects to mimic `realistic' images expected from an SKA-like observation to ascertain whether the WST remains sensitive to the cosmological signal. Finally, we perform a Fisher Matrix analysis to quantitatively assess the performance of the WST applied to 21-cm images. 

The remainder of this paper is organised as follows. In Section~\ref{sec:Method} we summarise the 21-cm simulations used in this work along with a description of the various astrophysical models. We then introduce the WST in Section~\ref{sec:Understanding} and apply it first to images of the raw, simulated 21-cm before considering it applied to realistic mock 21-cm images with the SKA. In Section~\ref{sec:forecasts} we perform Fisher Matrix forecasts for the WST applied to mock 21-cm images from an SKA-like observation. Finally, in Section~\ref{sec:conclusion} we provide a brief discussion before finishing with our closing remarks. Unless stated otherwise, all quantities are in in co-moving units and we adopt the cosmological parameters:  ($\Omega_\Lambda$, $\Omega_{\rm M}$, $\Omega_b$, $n$, $\sigma_8$, $H_0$) = (0.69, 0.31, 0.048, 0.97, 0.81, 68 km s$^{-1}$ Mpc$^{-1}$), consistent with recent results from the Planck mission \citep{Planck:2020}.

\section{Simulating the 21-cm signal} \label{sec:Method}

In this work, we simulate the cosmic 21-cm signal using the semi-numerical simulation code \cmfst{}\footnote{https://github.com/21cmfast/21cmFAST}\citep{Mesinger:2007p122,Mesinger:2011p1123}. In particular, we use the latest public release v3 \citep{Murray:2020} and the \citet{Park:2019} astrophysical parameterisation which provides prescriptions for the ultra-violet (UV) and X-ray properties of the galaxies responsible for reionisation. Below, we briefly summarise the key ingredients for simulating the cosmic 21-cm signal with \cmfst{}, and defer the interested reader to the aforementioned publications for further details.

\subsection{Galaxy UV properties}

First, we assume that the stellar mass, $M_{\ast}$, of a typical galaxy is directly related to its host halo mass, $M_{\rm h}$ \citep[e.g.][]{Kuhlen:2012p1506,Dayal:2014b,Behroozi:2015p1,Mitra:2015,Mutch:2016,Ocvirk:2016,Sun:2016p8225,Yue:2016,Hutter:2020},
\begin{eqnarray} \label{}
M_{\ast}(M_{\rm h}) = f_{\ast}\left(\frac{\Omega_{\rm b}}{\Omega_{\rm m}}\right)M_{\rm h},
\end{eqnarray}
and that the fraction of galactic gas in stars, $f_{\ast}$, can also be related to the host halo mass through a power-law relation\footnote{
This power-law behaviour between $M_{\ast}$ and $M_{\rm h}$ is consistent with the mean behaviour recovered from both semi-empirical fits to observations \citep[e.g.][]{Harikane:2016,Tacchella:2018,Behroozi:2019} and semi-analytic model predictions \citep[e.g][]{Mutch:2016,Yung:2019,Hutter:2020}.
},
\begin{eqnarray} \label{}
f_{\ast} = f_{\ast, 10}\left(\frac{M_{\rm h}}{10^{10}\,M_{\odot}}\right)^{\alpha_{\ast}}.
\end{eqnarray}
Here, $\alpha_{\ast}$ is the power-law index and $f_{\ast, 10}$ is the normalisation of this expression evaluated for a dark matter halo of mass $10^{10}$~$M_{\odot}$, and both are free parameters in this model.

The star-formation rate (SFR) for these galaxies is then determined by dividing the stellar mass by a characteristic time-scale, 
\begin{eqnarray} \label{}
\dot{M}_{\ast}(M_{\rm h},z) = \frac{M_{\ast}}{t_{\ast}H^{-1}(z)},
\end{eqnarray}
where $H^{-1}(z)$ is the Hubble time and $t_{\ast}$ is a free parameter allowed to vary between zero and unity.

Similar to above, the escape fraction of UV ionising photons, $f_{\rm esc}$, is assumed to be a power-law with halo mass
\begin{eqnarray} \label{}
f_{\rm esc} = f_{\rm esc, 10}\left(\frac{M_{\rm h}}{10^{10}\,M_{\odot}}\right)^{\alpha_{\rm esc}},
\end{eqnarray}
with power-law index, $\alpha_{\rm esc}$ and the normalisation $f_{\rm esc, 10}$ both being free parameters.

Finally, only some fraction of halos contribute to reionisation due to inefficient cooling and/or feedback processes in low mass halos which prevent the formation of star-forming galaxies. To account for this a duty-cycle is included,
\begin{eqnarray} \label{eq:duty}
f_{\rm duty} = {\rm exp}\left(-\frac{M_{\rm turn}}{M_{\rm h}}\right).
\end{eqnarray}
with the fraction, $(1 - f_{\rm duty})$, accounting for the suppression of star-forming galaxies below some scale $M_{\rm turn}$ \citep[e.g.][]{Shapiro:1994,Giroux:1994,Hui:1997,Barkana:2001p1634,Springel:2003p2176,Mesinger:2008,Okamoto:2008p2183,Sobacchi:2013p2189,Sobacchi:2013p2190} which is also free parameter.

One of the key advantages of this parameterisation is that galaxy UV luminosity functions (LFs), following some simple conversions, can be produced allowing the \cmfst{} parameterisation to be constrained by observed high-$z$ galaxy LFs.

\subsection{Galaxy X-ray properties}

It is thought that stellar remnants in the first galaxies emit X-rays capable of escaping the host galaxy and heating the IGM. This X-ray heating of the IGM is modelled by computing the cell-by-cell angle-averaged specific X-ray intensity, $J(\boldsymbol{x}, E, z)$, (in erg s$^{-1}$ keV$^{-1}$ cm$^{-2}$ sr$^{-1}$),
\begin{equation} \label{eq:Jave}
J(\boldsymbol{x}, E, z) = \frac{(1+z)^3}{4\pi} \int_{z}^{\infty} dz' \frac{c dt}{dz'} \epsilon_{\rm X}  e^{-\tau}.
\end{equation}
This is calculated by integrating the co-moving X-ray specific emissivity, $\epsilon_{\rm X}(\boldsymbol{x}, E_e, z')$ back along the light-cone accounting for attenuation of the X-rays by the IGM given by $e^{-\tau}$. The specific emitted emissivity, $E_{\rm e} = E(1 + z')/(1 + z)$, is then,
\begin{equation} \label{eq:emissivity}
\epsilon_{\rm X}(\boldsymbol{x}, E_{\rm e}, z') = \frac{L_{\rm X}}{\rm SFR} \left[ (1+\bar{\delta}_{\rm nl}) \int^{\infty}_{0}{\rm d}M_{\rm h} \frac{{\rm d}n}{{\rm d}M_{\rm h}}f_{\rm duty} \dot{M}_{\ast}\right],
\end{equation}
where the quantity in square brackets is the SFR density along the light-cone with $\frac{{\rm d}n}{{\rm d}M_{\rm h}}$ corresponding to the halo mass function (HMF)\footnote{Here, we adopt the Sheth-Tormen HMF \citep{Sheth:1999p2053}.
} and $\bar{\delta}_{\rm nl}$ is the mean, non-linear density in a shell centred on the simulation cell $(\boldsymbol{x}, z)$.

The emissivity is normalised by the specific X-ray luminosity per unit star formation escaping the host galaxies, $L_{\rm X}/{\rm SFR}$ (erg s$^{-1}$ keV$^{-1}$ $M^{-1}_{\odot}$ yr) which is assumed to follow a power-law with respect to photon energy, $L_{\rm X} \propto E^{- \alpha_X}$. This is then normalisation to the integrated soft-band ($<2$~keV) luminosity per SFR (in erg s$^{-1}$ $M^{-1}_{\odot}$ yr),
\begin{equation} \label{eq:normL}
  L_{{\rm X}<2\,{\rm keV}}/{\rm SFR} = \int^{2\,{\rm keV}}_{E_{0}} dE_e ~ L_{\rm X}/{\rm SFR} ~,
\end{equation}
where $E_0$ corresponds to the energy threshold below which X-ray photons are absorbed by the host galaxy. This amounts to three free parameters for the X-ray properties, $L_{{\rm X}<2\,{\rm keV}}/{\rm SFR}$, $E_0$ and $\alpha_X$, where for this work we assume $\alpha_X = 1$ consistent high-mass X-ray binary observations in the local Universe \citep{Mineo:2012p6282,Fragos:2013p6529,Pacucci:2014p4323}.

\subsection{Ionisation and Thermal State of the IGM}

\cmfst{} generates the velocity and evolved density fields using second-order Lagrange perturbation theory \citep[e.g][]{Scoccimarro:1998p7939}. The ionisation field is then obtained from the evolved density field using an excursion-set approach \citep[e.g.][]{Furlanetto:2004p123}. Here, the cumulative number of ionising photons, $n_{\rm ion}$, are compared to the total number of neutral hydrogen atoms plus cumulative recombinations, $\bar{n}_{\rm rec}$ \citep[e.g.][]{Sobacchi:2014p1157} in spheres of decreasing radii. A simulation cell is ionised when,
\begin{eqnarray} \label{eq:ioncrit}
n_{\rm ion}(\boldsymbol{x}, z | R, \delta_{R}) \geq (1 + \bar{n}_{\rm rec})(1-\bar{x}_{e}),
\end{eqnarray}
where the $(1-\bar{x}_{e})$ factor corresponds to ionisation by X-rays. The cumulative number of ionising photons per baryon inside a spherical region of size, $R$ and overdensity, $\delta_{R}$ is,
\begin{eqnarray} \label{eq:ioncrit2}
n_{\rm ion} = \bar{\rho}^{-1}_b\int^{\infty}_{0}{\rm d}M_{\rm h} \frac{{\rm d}n(M_{h}, z | R, \delta_{R})}{{\rm d}M_{\rm h}}f_{\rm duty} \dot{M}_{\ast}f_{\rm esc}N_{\gamma/b},
\end{eqnarray}
where $\bar{\rho}_b$ is the mean baryon density and the total number of ionising photons per stellar baryon is given by $N_{\gamma/b}$\footnote{By default \cmfst{} assumes $N_{\gamma/b}=5000$, consistent with that of a Salpeter initial mass function \citep{Salpeter:1955}.}

The thermal state of the neutral IGM is then determined by self-consistently calculating the heating and cooling rates from structure formation, Compton scattering off CMB photons and heating following partial ionisations as well as X-ray heating and ionisations. This sets the IGM spin temperature, $T_{\rm S}$, which is obtained by the weighted mean of the gas, $T_{\rm K}$, and CMB, $T_{\rm CMB}$, temperatures,
\begin{eqnarray} \label{}
T^{-1}_{\rm S} = \frac{T^{-1}_{\rm CMB} + x_{\alpha}T^{-1}_{\alpha} + x_{\rm c}T^{-1}_{\rm K}}{1 + x_{\alpha} + x_{\rm c}},
\end{eqnarray}
and depends on both the local gas density and the intensity of the Lyman-$\alpha$ (Ly$\alpha$) radiation impinging on the simulation cell. Here, $x_{\alpha}$ is the Wouthuysen-Field coupling coefficient \citep{Wouthuysen:1952p4321,Field:1958p1} and $x_{\rm c}$ is the collisional coupling coefficient between the free electrons and CMB photons. The Ly$\alpha$ background is formed by the contribution of X-ray excitations of neutral hydrogen atoms and the direct stellar emission of Lyman band photons by the first sources. For further details, see \citet{Mesinger:2011p1123}.

\subsection{21-cm Brightness Temperature}

The cosmic 21-cm signal is measured as a brightness temperature fluctuation relative to the background CMB, \citep[e.g.][]{Furlanetto:2006p209}:
\begin{eqnarray} \label{eq:21cmTb}
\delta T_{\rm b}(\nu) &=& \frac{T_{\rm S} - T_{\rm CMB}(z)}{1+z}\left(1 - {\rm e}^{-\tau_{\nu_{0}}}\right)~{\rm mK},
\end{eqnarray}
where $\tau_{\nu_{0}}$ is the optical depth of the 21-cm line,
\begin{eqnarray}
\tau_{\nu_{0}} &\propto& (1+\delta_{\rm nl})(1+z)^{3/2}\frac{x_{\hi{}}}{T_{\rm S}}\left(\frac{H}{{\rm d}v_{\rm r}/{\rm d}r+H}\right).
\end{eqnarray}
This expression depends on the neutral hydrogen fraction, $x_{\hi{}}$, gas overdensity, $\delta_{\rm nl} \equiv \rho/\bar{\rho} - 1$, the Hubble parameter, $H(z)$, and the line-of-sight gradient of the peculiar velocity. Here, everything is evaluated at the redshift $z = \nu_{0}/\nu - 1$ and we have dropped the spatial dependence for brevity.

\begin{table*}
\begin{tabular}{@{}lccccccccc}
\hline
Model Type  & ${\rm log_{10}}(f_{\ast,10})$ & $\alpha_{\ast}$ & ${\rm log_{10}}(f_{\rm esc,10})$ & $\alpha_{\rm esc}$ & $t_{\ast}$ & ${\rm log_{10}}(M_{\rm turn})$ & ${\rm log_{10}}\left(\frac{L_{{\rm X}<2{\rm keV}}}{\rm SFR}\right)$ & $E_0$   \\
               &  &  &  &  & & $[{\rm M_{\sun}}]$ & $[{\rm erg\,s^{-1}\,M_{\sun}^{-1}\,yr}]$ &  $[{\rm keV}]$ \\
\hline
\vspace{0.8mm}
Fiducial Model & $-1.30$ & $0.50$ & $-1.00$ & $-0.50$ & $0.5$ & $8.7$ & $40.50$ &  $0.50$\\
\hline
\vspace{0.8mm}
Cold Reionisation &  $-1.30$ & $0.50$ & $-1.00$ & $-0.50$ & $0.5$ & $8.7$ & $38.00$ &  $0.50$  \\
\hline
\vspace{0.8mm}
Large Haloes &   $-0.70$ & $0.50$ & $-1.00$ & $-0.50$ & $0.5$ & $9.9$ & $40.50$ &  $0.50$  \\
\hline
\vspace{0.8mm}
Extended Reionisation &  $-1.65$ & $0.50$ & $-1.00$ & $-0.50$ & $0.5$ & $8.0$ & $40.50$ &  $0.50$ \\
\hline
\end{tabular}
\caption{A summary of the astrophysical parameters used for our four different reionisation models used for exploring the WST applied to the 21-cm signal. See Section~\ref{sec:setup} for a description of each model.}
\label{tab:Models}
\end{table*} 

\subsection{Simulation Setup} \label{sec:setup}

To gain valuable insights into how the WST extracts astrophysical information from the cosmic 21-cm signal we consider four different reionisation scenarios. For each, we simulate the full 21-cm light cone spanning $z=5.9 - 35$ using \cmfst{} with a transverse comoving length of 250 Mpc and 128 voxels per side length (see Appendix~\ref{app:convergence} for discussions on convergence testing). In Table~\ref{tab:Models} we provide the astrophysical parameters for each of these four models, and provide a brief description of each below:
\begin{itemize}
\item[1.] \textit{Fiducial Model}: In this work, we adopt the default model considered in \citet{Park:2019} as our fiducial model\footnote{Though see \citet{Qin:2021} for a more up-to-date model using a broader range of recent observations.}. This was found to match a range of observational constraints including observed UV LFs at $z=6-10$ and the electron scattering optical depth, $\tau_{\rm e}$.\\
\item[2.] \textit{Cold Reionisation}: Here, reionisation proceeds as per our fiducial model above except we significantly increase the amplitude of the 21-cm signal by considering a scenario whereby the IGM undergoes very little or no X-ray heating. This class of model is referred to as `cold' reionisation, \citep[e.g.][]{Mesinger:2014p244,Parsons:2014p781}. \\
\item[3.] \textit{Large Haloes}: We increase our characteristic turn-over scale for star-forming galaxies, $M_{\rm turn}$, to ensure reionisation is driven by larger, more biased galaxies leading to a different ionisation morphology than our fiducial model. \\
\item[4.] \textit{Extended Reionisation}: We decrease the ionising efficiency of our star-forming galaxies to produce a slower, more extended reionisation driven by a dominant population of faint star-forming galaxies. This results in a considerably later reionisation ($z\sim5$) while also further reducing the amplitude of the 21-cm signal.
\end{itemize}

The models above are primarily chosen to be illustrative of different features present in the 21-cm signal, not necessarily to be consistent with existing observational constraints. The resultant reionisation history (IGM neutral fraction as a function of redshift) and the mean 21-cm brightness temperature signal (global signal) are shown in the top two panels of Figure~\ref{fig:Global_vs_S0}.

\section{Applying the Wavelet Scattering Transform to the 21-cm Signal} \label{sec:Understanding}

\subsection{The Wavelet Scattering Transform} \label{sec:WST}

The Wavelet Scattering Transform (WST) is the convolution of a target field (image), $I(\boldsymbol{x})$, by a family of rotated and dilated wavelet filters. The key feature of wavelet filters is that they preserve locality both spatially and in frequency (unlike the Fourier Transform). Following the convolution operation, the modulus of the filtered field is calculated before spatially averaging to compress the information down to a single scattering coefficient. In this work, we specifically focus on applying the WST to 2D images, following closely the approach used by \citet{Cheng:2020} which explored the WST for analysing the weak lensing signal in 2D images. In particular, we adopt the Morlet filter as our wavelet family. Note however, that the approach can be generalised to any dimension. It will be interesting to determine how well the WST performs in 3D with the 21-cm signal having a clear redshift dependence along the line-of-sight. However, initial exploration using wavelets by \citet{Trott:2016p7921} highlighted the improvements in handling this redshift dependence relative to the power spectrum.

Given an input 2D image of the 21-cm signal, $I_{0}(x,y)$, the zeroth order scattering coefficient is simply the mean of the signal,
\begin{eqnarray} 
s_{0} = \langle I_{0}(x,y)\rangle,
\end{eqnarray} 
which, for the 21-cm signal, is simply the mean brightness temperature, $\bar{\delta T}_{\rm b}$, that is, the global 21-cm signal.

The family of first-order fields, $I_{1}(x,y)$, are obtained by the convolution of the original image by a family of wavelets, $\psi^{j_1,l_1}(x,y)$, where $j$ and $l$ correspond to different physical extents and rotations of the filters. Taking the modulus and then averaging we obtain the family of first-order scattering coefficients, $s^{\,j_1,l_1}_{1}$,
\begin{eqnarray} 
s^{\,j_1,l_1}_{1} = \langle \, | I_{0} \ast \psi^{j_1,l_1} \, | \rangle.
\end{eqnarray} 

We can then extend this to second-order by filtering our first-order fields by another set of filters, $\psi^{j_2,l_2}(x,y)$. Taking the modulus and averaging following the convolution of this second family of filters produces the second-order scattering coefficients, $s^{\,j_1,l_1,j_2,l_2}_{2}$,
\begin{eqnarray} 
s^{\,j_1,l_1,j_2,l_2}_{2} = \langle \, | I_{1} \ast \psi^{j_2,l_2} \, | \rangle = \langle \, | \, | I_{0} \ast \psi^{j_1,l_1} \,| \ast \psi^{j_2,l_2} \, | \rangle.
\end{eqnarray}
The key step above is the application of the modulus operation. Taking the modulus measures the strength of the spatial features prior to the secondary convolution and modulus operation. This measures the strength of the clustering of spatial features, which in effect is the non-Gaussian information. This same procedure can be followed to produce higher-order coefficients, however, in this work we only consider the scattering coefficients up to the second order. By design, the information cascades very quickly, thus going to even higher order coefficients usually does not capture much additional information beyond the second order. The largest physical scale possible (denoted $J$) is set by the requirement that $2^J$ cannot exceed the number of pixels in any one dimension of the image. The total number of available scattering coefficients is dictated by the number of filters used with $J$ defining the maximal spatial scale and $L$, the total number of possible orientations (i.e. rotations) of the filters (each is rotated by $\pi/L$). Thus, with a total of $J\times L$ filters used there are $J^n\,L^n$ possible combinations of coefficients, with $n$ being the order of scattering coefficients we are considering (i.e. in our case, $n=2$).

In this work, we reduce the total number of scattering coefficients by averaging over all possible orientations of the filters. This assumes that the 21-cm signal is isotropic. That is, for the 21-cm signal we measure the following,
\begin{eqnarray} 
S_{0} &=& s_{0}, \\
S^{\,j_1}_{1} &=& \langle s^{\,j_1,l_1}_{1} \rangle_{l_1}, \\
S^{\,j_1,j_2}_{2} &=& \langle s^{\,j_1,l_1,j_2,l_2}_{2} \rangle_{l_1,l_2}.
\end{eqnarray} 
This reduces the total number of scattering coefficients down to a more manageable number of $1+J+J^{2}$. For our simulation setup (128 voxels per side length), this corresponds to $J=7$ and for convenience we only consider four distinct rotations, $L=4$.  The physical scales correspond to a dyadic sequence ($2^{j}$) resulting in 7 logarithmically spaced bins ranging from $\sim 5.2$~Mpc ($j=0$) to $\sim 333.3$~Mpc ($j=6$). To obtain the scattering coefficients using the WST we use the convenient \textsc{\small python} package \textsc{\small kymatio}\footnote{https://www.kymat.io/}\citep{Andreux:2018}. For the second-order scattering coefficients \textsc{\small kymatio} by default only calculates those with $j_{2} > j_{1}$ as the coefficients with $j_{2} \leq j_{1}$ contain very little cosmological information as the `high' frequency information (small physical scales) gets discarded first. Conveniently, this further reduces the total number of scattering coefficients down to $1 + 7 + 21 = 29$.

\begin{figure*} 
	\begin{center}
		\includegraphics[trim = 0.2cm 0.6cm 0cm 0.5cm, scale = 0.5]{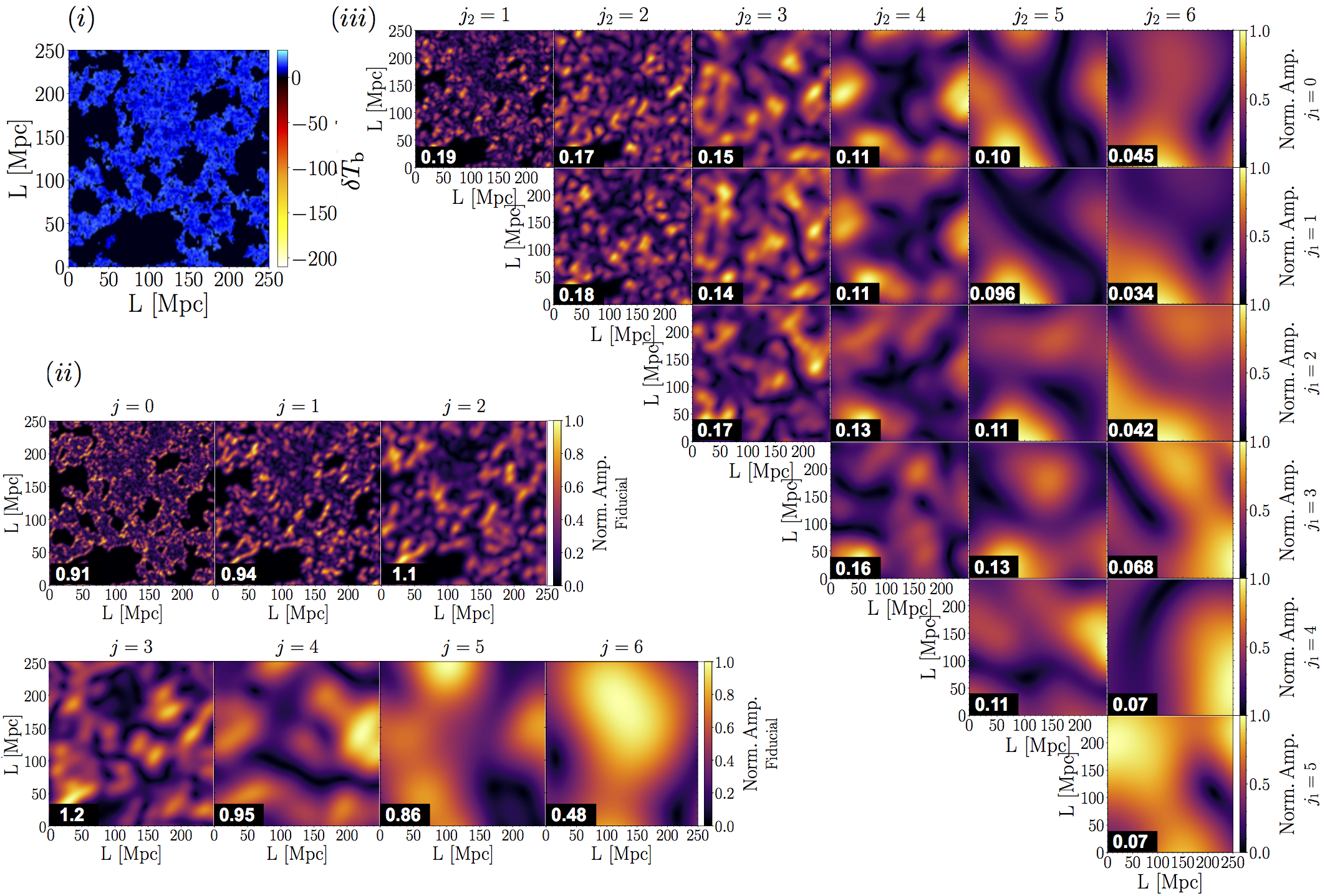}
	\end{center}
\caption[]{The wavelet scattering transform applied to a single, 21-cm image at $z = 7.5$ ($\bar{x}_{\hi{}} \sim 0.5$) from our fiducial model. (\textit{i}): The simulated 21-cm brightness temperature output directly from \cmfst{}, (\textit{ii}): The 21-cm signal filtered by the first-order family of wavelets in increasing order of spatial scale, $j$, (\textit{iii}): The 21-cm signal filtered at second order, with the two scales represented by $j_{1}$ (rows) and $j_{2}$ (columns). Each filtered image is normalised to unity, with the white number in the bottom left corner corresponding to the scattering coefficient prior to normalisation (e.g. $S^{\,j}_{1}$ for first-order and $S^{\,j_{1}, j_{2}}_{2}$ for second-order). For all, the filtered images assume $l=2$ for the orientation.}
\label{fig:FilteredImage}
\end{figure*}

\subsection{Filtering a single 21-cm image} \label{sec:singleimage}

As a first demonstration of the WST to the 21-cm signal, we apply the WST to a single 21-cm image from one of our four simulations. In Figure~\ref{fig:FilteredImage} we show the 21-cm image at $z=7.5$ ($\bar{x}_{\hi{}} \sim 0.5$) from our fiducial simulation (panel (\textit{i})), and the first and second-order filtered images for $l=2$ (panels (\textit{ii}) and (\textit{iii}), respectively) along with the corresponding scattering coefficients in the bottom left corner of each panel (white text). Each filtered image is normalised to unity to amplify the spatial features rather than the amplitude of the signal (the scattering coefficients are calculated prior to the normalisation).

We can see from Figure~\ref{fig:FilteredImage} that as we increase $j$ we are progressively becoming more sensitive to physically larger-scale features in the 21-cm signal. This is evident as for $j<4$, the original ionised regions (i.e. ionised bubbles) are still clearly present for the first-order filtered images. The second-order filters then pick up the clustering of these ionised bubbles of this same scale. Eventually, for increasing $j$ we begin to consider scales larger than the individual ionised regions, becoming more sensitive to large-scale fluctuations in brightness temperature, driven primarily by thermal effects from X-ray heating. Importantly, this interpretation is only illustrative, as it is difficult to interpret the underlying physics of reionisation from a single image. It is more illuminating to follow the evolution of the scattering coefficients as a function of redshift (frequency), which we investigate next.

\subsection{WST of the raw, simulated 21-cm light-cone} \label{sec:S-evolution}

To truly understand what we are learning about the EoR and CD from the WST, we switch from focusing on a single 21-cm image from one simulation to the full 21-cm light cone from the four different reionisation simulations we discussed in Section~\ref{sec:setup}. In doing so, we can more readily compare the performance of the WST against our established understanding of how we know the 21-cm signal behaves. For this, we apply the WST to images extracted from our simulated 21-cm light cone with a cadence of 1~MHz between 50 - 206 MHz (corresponding to $z = 27.4 - 5.9$). The lower end of this range (50 MHz) corresponds to the low frequency edge of the SKA observing band, while the upper end is dictated by the end of reionisation for our fiducial setup. For each image, we then calculate all relevant scattering coefficients.

\subsubsection{$S_{0}$ coefficients}

\begin{figure} 
	\begin{center}
		\includegraphics[trim = 0.2cm 0.6cm 0cm 0.5cm, scale = 0.53]{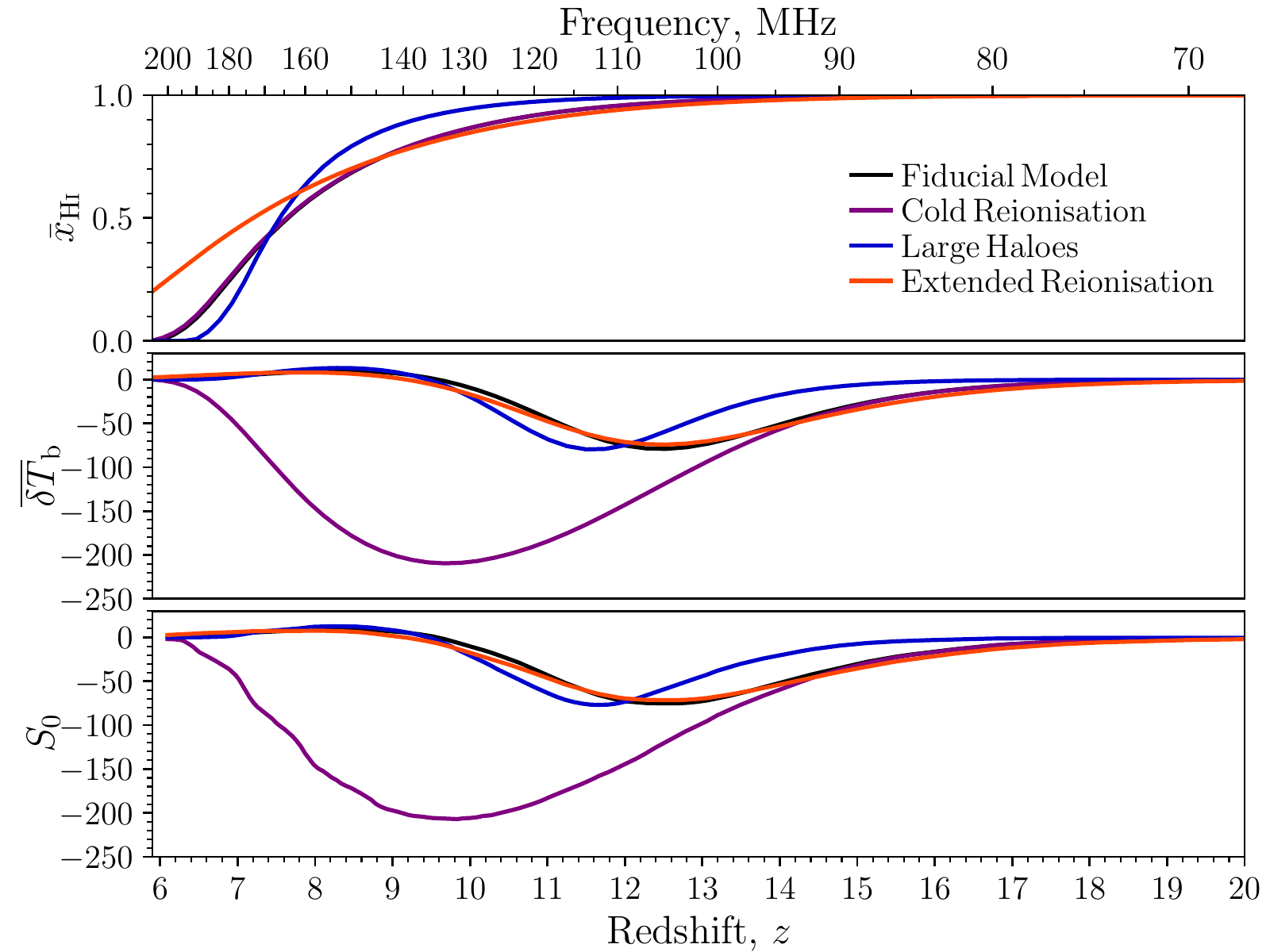}
	\end{center}
\caption[]{The globally averaged IGM neutral fraction, $\bar{x}_{\hi{}}$ (top panel), and the mean 21-cm brightness temperature, $\overline{\delta T_{\rm b}}$ (middle panel) as a function of redshift calculated from the four different astrophysical models outlined in Section~\ref{sec:setup}. For comparison, we provide the evolution in the $S_{0}$ scattering coefficient recovered from the Wavelet Scattering Transform in the bottom panel.}
\label{fig:Global_vs_S0}
\end{figure}

In Figure~\ref{fig:Global_vs_S0} we compare the evolution of the IGM neutral fraction, $\bar{x}_{\hi{}}$, and the mean 21-cm brightness temperature, $\overline{\delta T_{\rm b}}$ against the $S_{0}$ coefficient for all four of our astrophysical models. Comparing the second and third panels, it is clear that the $S_{0}$ coefficient is equivalent to the mean 21-cm brightness temperature, which was to be expected from Section~\ref{sec:WST}. Note, while the evolution of the $S_{0}$ coefficient appears to be less smooth than the mean brightness temperature, this is because they are computed differently. For $\overline{\delta T_{\rm b}}$ it is measured from each of the full 3D co-eval (single redshift) cubes used to construct the light cone. Whereas for $S_{0}$ we are determining it from a single 2D image. Thus for $S_{0}$ we are calculating the mean signal from considerably less information resulting in considerably more variation between images. Importantly for this work, a 21-cm interferometer experiment only measures the statistical fluctuations in the 21-cm signal, not the mean signal. Thus in practice, $S_{0}$ extracted from real 21-cm images will return a mean zero signal. This mean signal is what is observed by global signal experiments, such as the Experiment to Detect the Global EoR Signature (EDGES; \citealt{Bowman:2010p6724}).

\subsubsection{$S_{1}$ coefficients} \label{sec:s1}

As discussed in-depth by \citet{Cheng:2020}, the $S_{1}$ coefficients can be closely related to the power spectrum. For example, for the 21-cm signal, the power spectrum, $P_{21}(\boldsymbol{k})$ is calculated by taking the square of the Fourier transformed fluctuations in the 21-cm brightness temperature, $P_{21}(\boldsymbol{k}) \propto \langle | \delta_{21}(\boldsymbol{k}) |^{2} \rangle$ where $\delta_{21}(\boldsymbol{k}) \equiv \delta T_{\rm b}(\boldsymbol{x})/\overline{\delta T_{\rm b}}(\boldsymbol{x})-1$. If we instead substitute the Morlet wavelet with a plane-wave (i.e. Fourier) filter, $\psi = {\rm e}^{-i \boldsymbol{k.x}}$, and take the square of the resultant filtered image we obtain,
\begin{eqnarray} 
P(\boldsymbol{k}) \propto \langle \, | I_{1} \, |^{2} \rangle \propto \langle \, | I_{0} \ast {\rm e}^{-i \boldsymbol{k.x}} \, |^{2} \rangle.
\end{eqnarray} 
Remembering that the first-order scattering coefficients are given by $s^{\,j,l}_{1} = \langle \, | I_{0} \ast \psi^{j,l} \, | \rangle$ it is straightforward to see the similarities between the scattering coefficients and the Fourier power spectrum. Both measure the strength of the spatial fluctuations as a function of scale. Therefore, each of the $S_{1}$ scattering coefficients can be interpreted as coarsely binned power spectra. The key advantage of the WST is that by convolving the signal with a Morlet filter, the filtered signal retains both the structural and phase information in both real/redshift space and in wavelet space, as is evident in Figure~\ref{fig:FilteredImage}. In contrast, the Fourier transform de-localises the signal. Subsequently, by convolving the wavelet filtered images iteratively with the same family of filters then enables the exploration of the clustering of the phase information (i.e. the second-order coefficients, discussed further in the next section).

\begin{figure*} 
	\begin{center}
	 	\includegraphics[trim = 0.4cm 0.5cm 0cm 0.3cm, scale = 0.89]{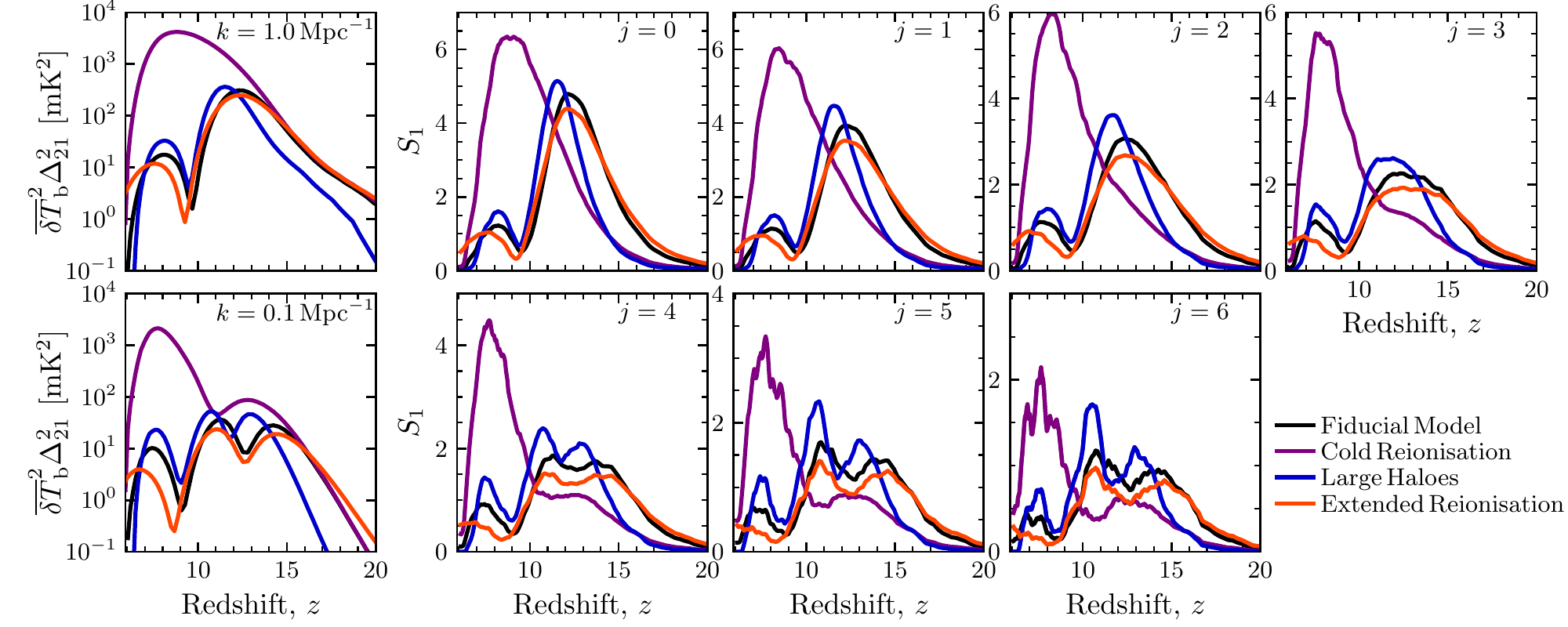}
	\end{center}
\caption[]{A direct comparison between the evolution of the 21-cm power spectrum and the recovered $S_{1}$ scattering coefficients. Left most column is the evolution in the 21-cm power ($k=1.0$ and 0.1~Mpc$^{-1}$, respectively) for the four different astrophysical models. In the remaining panels, we present the evolution of all possible $S_{1}$ scattering coefficients. Note that for the Cold Reionisation model, we reduce the amplitude of these coefficients by a factor of 3 to more easily present the evolution of all four models.}
\label{fig:21cmPS_vs_S1}
\end{figure*}

It is well established that the evolution of the 21-cm power as a function of redshift shows three distinct peaks at large scales (i.e. $k = 0.1$~Mpc$^{-1}$, see e.g. \citealt{Pritchard:2007p3787,Baek:2010p6357}). These correspond to three important epochs; the EoR, the Epoch of Heating (EoH) and Wouthuysen-Field (WF) coupling \citep{Wouthuysen:1952p4321,Field:1958p1} in increasing order of redshift. In the left most panel of Figure~\ref{fig:21cmPS_vs_S1}, we show the 21-cm power at two different spatial scales ($k = 1.0$ and 0.1~Mpc$^{-1}$, respectively) for the four astrophysical models outlined in Section~\ref{sec:setup}. Across these four models, the location of these peaks differ in both amplitude and redshift, dependent on the specifics of the astrophysics in each model. For example, the cold reionisation model is $\sim2$ orders of magnitude larger than the other models, highlighting the impact that limited IGM heating can have on the amplitude of the 21-cm brightness temperature signal. Further, owing to the limited X-ray heating there is no EoH and thus there are only two peaks.

In the remaining panels of Figure~\ref{fig:21cmPS_vs_S1} we present the evolution of the $S_{1}$ scattering coefficients for the four different astrophysical models, in increasing order of physical scale ($j$) for the wavelet filters. As discussed in Section~\ref{sec:singleimage} the different $j$ wavelet filters access different spatial scales, with low $j$ corresponding to small physical scales and high $j$ corresponding to large physical scales. Thus, the evolution of the $S_{1}$ scattering coefficients mirror the equivalent evolution of the 21-cm power at different Fourier scales. Low $j$ scales are equivalent to large $k$ Fourier modes, which is clearly evident as the evolution of the $k=1.0$~Mpc$^{-1}$ mode (dashed curves) closely matches the low $j$ scattering coefficient evolution. Equally the large $j$ coefficients more closely resemble that of the smaller scale $k=0.1$~Mpc$^{-1}$ Fourier mode, showing the distinctive three peaked structure for the cosmic 21-cm signal. Thus, the first-order scattering coefficients comparable information to the equivalent Fourier power spectrum.

It is important to note that for Figure~\ref{fig:21cmPS_vs_S1} and for the remainder of this work, we perform a simple box-car filtering to smooth out noise features for the evolution of the scattering coefficients. In effect, we perform a running average of each scattering coefficient over a 10 MHz range (we apply the WST on images at a 1 MHz cadence). In Appendix~\ref{app:Smoothing} we discuss this in more detail, highlighting the smoothed coefficients relative to the raw output (at 1 MHz cadence) in Figure~\ref{fig:Smoothing}. Note, these noise like features could be reduced by considering images of larger physical extent.

\subsubsection{$S_{2}$ coefficients} \label{sec:S2}

\begin{figure*} 
	\begin{center}
		\includegraphics[trim = 1cm 2cm 0cm 0.5cm, scale = 0.46]{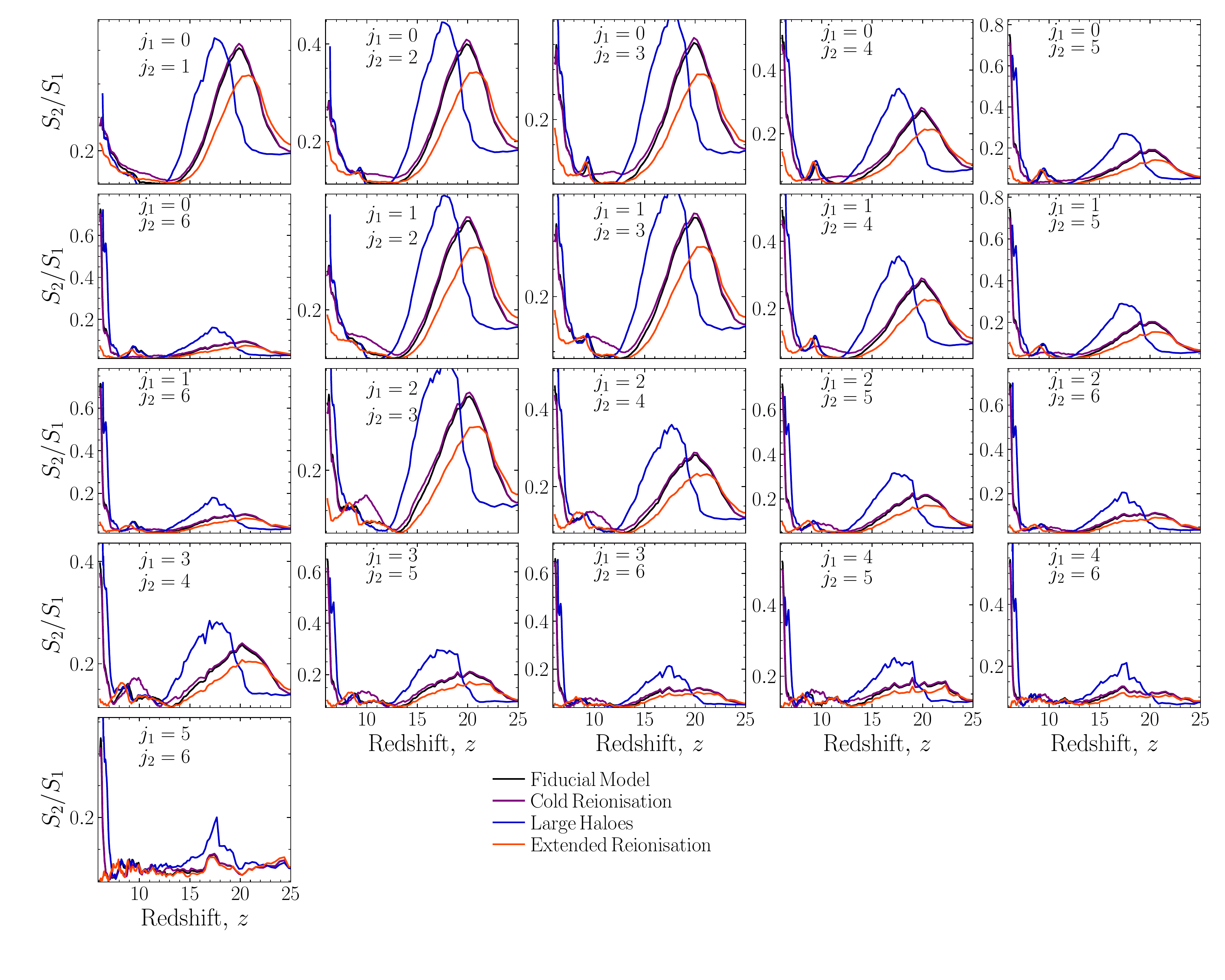}
	\end{center}
\caption[]{A comparison of the evolution for all combinations of the de-correlated $S_{2}$ coefficients (dividing out the dependence from the $S_{1}$ coefficients) for the four astrophysical models outlined in Section~\ref{sec:setup}.}
\label{fig:S2Coefficients}
\end{figure*}

In the previous section, we highlighted the similarities between the $S_{1}$ coefficients and the Fourier power spectrum. Additionally, we highlighted the key advantage of the WST of the Fourier transform in that it always preserves both the structural and phase information. Following the WST properties outlined in Section~\ref{sec:WST}, the $S_{2}$ coefficients are obtained from the subsequent convolution of the first-order family of filtered images after taking the modulus. If the $S_{1}$ coefficients are related to the Fourier power spectrum (i.e the spatial information of the original input field), the $S_{2}$ coefficients are then related to the power spectrum of the first-order filtered images of the input field. Thus, the $S_{2}$ coefficients measure the spatial information (clustering) of the spatial information of the original field (see \citealt{Cheng:2020} for further details). In effect, these $S_{2}$ coefficients contain information up to the 4-point function of the input signal field. Thus, in analogy to the $S_1$ coefficients, $S_{2}$ can be thought of as containing information from coarsely binned bispectra and trispectra (i.e. the Fourier transform of the three and four point correlation functions). The advantage of the WST is that this binning of the higher order coefficients is considerably less noisy than that of the higher order polyspectra\footnote{Secondly, the WST is considerably cheaper in terms of total compute time than any higher-order polyspectra as it does not require binning of the statistical information (only convolutions and modulus operations).}, the bispectrum and trispectrum \citep{Cheng:2021b}. However, unlike for the power spectrum above, we do not demonstrate the common features to these higher order functions. Primarily, the difficulty arises from determining the equivalence between the WST binning and the appropriate closure configuration for the bispectrum or trispectrum. Such a study is beyond the scope of this work.

Equally, via the same properties of the WST, the $S_{2}$ coefficients are strongly correlated to the $S_{1}$ coefficients (since they are the filtered version following the modulus of $S_{1}$). Thus, if we present the evolution of the $S_{2}$ coefficients (as we have done for the $S_{1}$ coefficients above) these will be difficult to visually distinguish from the $S_{1}$ coefficients. Instead, we can produce de-correlated $S_{2}$ coefficients by simply dividing our $S^{\,j_{1},j_{2}}_{2}$ coefficients by the corresponding first-order coefficient, $S^{\,j_{1}}_{1}$ \citep[see e.g.][]{Bruna:2015,Cheng:2020}.

In Figure~\ref{fig:S2Coefficients} we present the evolution of the de-correlated $S_{2}$ coefficients as a function of redshift for the four different astrophysical models. These de-correlated coefficients contain notably different structural information to those from the $S_{1}$ coefficients as they are picking up information from the higher-order moments, making it difficult to perform a similar qualitative analysis as we performed previously. Note, here we explicitly provide all possible combinations to highlight the available information. However, in practice it is quite clear many of these combinations display very similar behaviour. As such, for the remainder of this work we will only present a select sub-set of all possible combinations for the de-correlated $S_{2}$ coefficients.

Primarily, these de-correlated $S_{2}$ coefficients obtain the largest amplitudes towards the tail-end of reionisation (lowest $z$), where the 21-cm brightness temperature is expected to be highly non-Gaussian (due to the complex morphology in the ionisation field). The peaks at the highest redshifts ($z\sim15-20$) are driven by the first appearance of the ionising sources (i.e. the rarest, most clustered peaks in the density field) and their imprint on the IGM brightness temperature through the WF-coupling effect. The signal is strongest on the smallest physical scales (low $j_{1}$ and/or low $j_{2}$) with the peak location shifting to lower redshift from the Extended reionisation model down to the Large Haloes model. The Extended reionisation model has the smallest $M_{\rm turn}$ (reionisation driven by smaller, fainter galaxies) meaning galaxies form first in this model, with the Large Haloes model (highest $M_{\rm turn}$) forming their first galaxies last. Thus, the additional information contained in the $S_{2}$ coefficients provides additional insight into the physical properties of the EoR and CD. 

It is important to re-emphasise that here we are presenting the de-correlated $S_{2}$ coefficients, not the $S_{2}$ coefficients. This choice is motivated by wanting to highlight the additional non-Gaussian cosmological information contained in these coefficients beyond that held in the $S_{1}$ coefficients. Finally, if the $S_{2}$ coefficients truly contain the non-Gaussian information describing the 21-cm signal, we would expect that the features presented here in Figure~\ref{fig:S2Coefficients} should disappear once we randomise the phases in our filtered first-order images of the 21-cm signal. In Appendix~\ref{app:Phases} we verify that this is indeed the case.

\subsection{The WST applied to realistic mock images of the 21-cm signal}

Previously, we have explored the WST applied directly to the raw simulated 21-cm signal. However, in practise from an interferometer experiment we will observe a corrupted 21-cm signal, which contains instrumental effects such as thermal noise and finite instrumental resolution. Thus, we now shift our attention to the WST applied to more realistic mock 21-cm images containing these effects to see if we are still sensitive to the 21-cm signal.

\subsubsection{Instrumental effects} \label{sec:instrument}

In this work, we adopt the SKA1--low as our fiducial 21-cm interferometer experiment. This choice is motivated by the expectation that the SKA will be able to provide three-dimensional tomographic images during the EoR. We model the SKA antenna configuration using the design outlined in the SKA System Baseline Design document\footnote{http://astronomers.skatelescope.org/wp-content/uploads/2016/09/SKA-TEL-SKO-0000422\textunderscore 02\textunderscore SKA1\textunderscore LowConfigurationCoordinates-1.pdf}. Here, 512 37.5m antennae stations are distributed within a 500m core radius and the system temperature, $T_{\rm sys}$, is modelled as $T_{\rm sys} = 1.1T_{\rm sky} + 40~{\rm K}$. In this work, we adopt $T_{\rm sky} = 60\left(\frac{\nu}{300~{\rm MHz}}\right)^{-2.55}~{\rm K}$ \citep{Thompson2007}. Images of the 21-cm signal are achieved by phase-tracking the sky, for which we assume a single six-hour track per night for a total observing time of 1000 hours.

To construct our mock images of the 21-cm signal using SKA1--low, we modify the publicly available \textsc{\small Python} module \textsc{\small 21cmSense}\footnote{https://github.com/jpober/21cmSense}\citep{Pober:2013p41,Pober:2014p35}. Nominally, \textsc{\small 21cmSense} uses the gridded $uv$-visibilities of any instrument design to yield an estimate of the thermal noise power-spectrum, $P_{\rm N}(k)$, given an input 21-cm power spectrum measurement according to the following expression,
\begin{eqnarray} \label{eq:NoisePS}
P_{\rm N}(k) \approx X^{2}Y\frac{\Omega^{\prime}}{2t}T^{2}_{\rm sys}.
\end{eqnarray} 
Here, $X^{2}Y$ is the cosmological conversion between observing bandwidth, frequency and co-moving distance, $\Omega^{\prime}$ is a beam-dependent factor derived in \citet{Parsons:2014p781}, $t$ is the total time spent by all baselines within a specific Fourier mode and $T_{\rm sys}$ is the system temperature. 

We model the effects of finite instrument resolution and thermal noise from our modified version of \textsc{\small 21cmSense} by performing the following steps:
\begin{itemize}
\item we first 2D Fourier transform the input (simulated) image of the 21-cm signal 
\item we then filter our image using the output gridded $uv$-visibilities for the SKA1--low baseline configuration from \textsc{\small 21cmSense}. Cells with finite $uv$-coverage are multiplied by unity, all others are set to zero
\item in each cell we then calculate the thermal noise power, $P_{\rm N}(k_{x}, k_{y})$, using Equation~\ref{eq:NoisePS} where $k_x$ and $k_y$ are the two transverse (on sky) directions and add a randomly sampled value from this measured power spectrum to each cell to mimic the effect of thermal noise
\item finally, we inverse Fourier transform back to image space to obtain our mock 21-cm image.
\end{itemize}

\subsubsection{Wedge mode removal} \label{sec:wedge}

Unfortunately, observations of the 21-cm signal are hit by another source of error. The visibilities ($uv$ coverage) of an interferometer experiment are frequency dependent. This means line-of-sight (frequency dependent) power can being thrown into the transverse (frequency independent) Fourier modes. This leakage of power results in a well-defined contaminated `wedge' in cylindrical 2D Fourier space \citep{Datta:2010p2792,Vedantham:2012p2801,Morales:2012p2828,Parsons:2012p2833,Trott:2012p2834,Thyagarajan:2013p2851,Liu:2014p3465,Liu:2014p3466,Thyagarajan:2015p7294,Thyagarajan:2015p7298,Pober:2016p7301,Murray:2018}. While it is possible to mitigate or `clean' these contaminated modes (see e.g.\ \citealt{Chapman:2019} for a comprehensive review, or by using machine learning \citealt{Gagnon-Hartman:2021}) in this work we take the conservative approach of cutting all contaminated `wedge' modes from our 21-cm light-cone.

The location of this `wedge' in 2D cylindrical space is determined by,
\begin{eqnarray} \label{eq:wedge}
k_{\parallel} =  mk_{\perp} + b
\end{eqnarray} 
where $k_{\parallel}$ and $k_{\perp}$ are the line-of-sight and transverse Fourier modes, $b$ is a additive buffer of $\Delta k_{\parallel} = 0.1 \,h$~Mpc$^{-1}$ extending beyond the horizon limit and,
\begin{eqnarray}
m = \frac{D_{\rm C}H_{0}E(z){\rm sin}(\theta)}{c(1+z)}.
\end{eqnarray} 
Here, $D_{\rm C}$ is the comoving distance, $H_{0}$ is the Hubble constant, $E(z) = \sqrt{\Omega_{\rm m}(1+z)^{3} + \Omega_{\Lambda}}$ and ${\rm sin}(\theta)$ accounts for the viewing angle of the telescope, for which we conservatively take as $\theta = \pi/2$ (i.e. a zenith pointing observation).

To mimic the removal of these `wedge' contaminated modes, prior to the steps outlined in Section~\ref{sec:instrument} we first:
\begin{itemize}
\item Construct a 3D volume of the 21-cm signal centred on our current observing frequency.
\item Fourier transform, apply the instrument effects and thermal noise as per the steps outlined in Section~\ref{sec:instrument}.
\item Excise all modes below the foreground contaminated `wedge' as defined by Equation~\ref{eq:wedge}.
\item {Inverse Fourier transform back.}
\end{itemize}

\subsubsection{Realistic mock 21-cm image}

\begin{figure*} 
	\begin{center}
		\includegraphics[trim = 0.2cm 0.6cm 0cm 0.5cm, scale = 0.65]{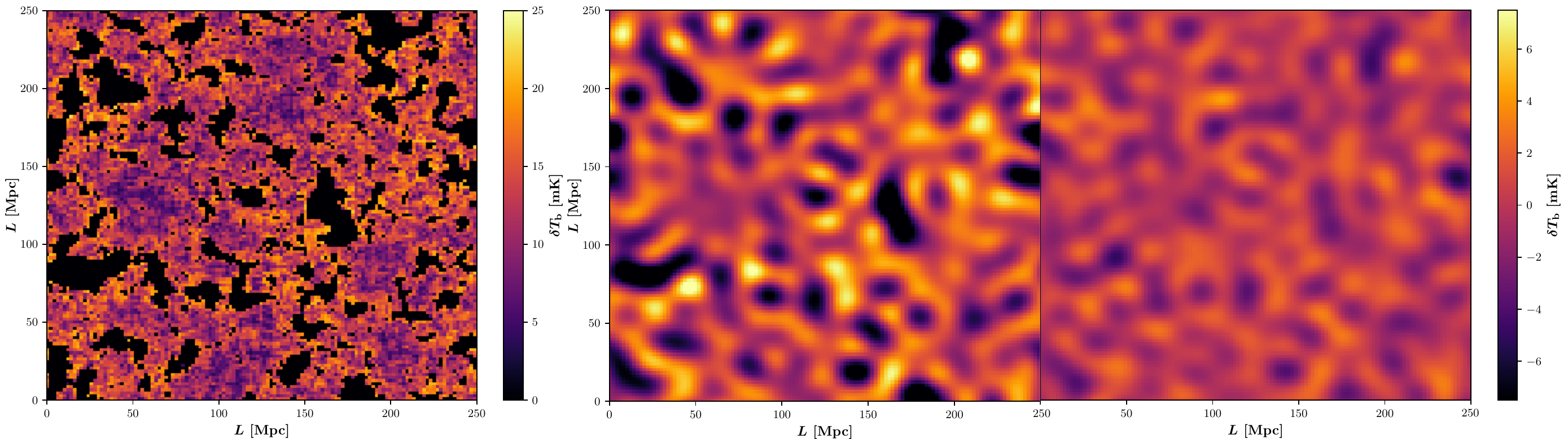}
	\end{center}
\caption[]{Schematic highlighting the incremental steps towards obtaining `realistic' mock 21-cm images from a 21-cm interferometer experiment such as SKA1--low, assuming a 1000 hour observation. \textit{Left panel:} the 21-cm signal extracted directly from our simulations, \textit{middle panel:} the 21-cm signal including thermal noise and finite resolution by applying the SKA $uv$-coverage (see Section~\ref{sec:instrument}) and in the \textit{right panel:} the 21-cm signal from the middle panel but with contaminated foreground `wedge' modes removed (see Section~\ref{sec:wedge}).}
\label{fig:ImageCreation}
\end{figure*}

In Figure~\ref{fig:ImageCreation} we provide a demonstration of this pipeline applied to a simulated 21-cm signal at 150~MHz ($z\sim8.5$) from our fiducial model. The left panel shows a 2D image of the raw, simulated 21-cm signal, the middle panel contains an idealised mock image following the steps outline in Section~\ref{sec:instrument} (e.g. perfect foreground removal) and in the right panel we show a realistic mock signal with the foreground `wedge' modes removed following Sections~\ref{sec:instrument} and~\ref{sec:wedge}.

Importantly, interferometer experiments do not measure the mean signal, instead they only measure the amplitude of the spatial fluctuations. In effect, we observe an image with the mean signal removed. Applying our pipelines above takes this into account (i.e. after Fourier transforming we remove the DC mode), thus, the mock images shown in the middle and left of Figure~\ref{fig:ImageCreation} only show the amplitude of the fluctuations around the mean. This has important consequences for the $S_{0}$ coefficients (i.e. the mean signal) since these will not be accessible from images of the 21-cm signal obtained from radio interferometers\footnote{We verified this by applying the WST to these mock images and recover $S_{0} = 0$ for all redshifts.}. The effective removal of the mean 21-cm signal from the image also impacts on the higher-order scattering coefficients. That is, their relative amplitudes reduce following mean removal.

The impact of finite resolution and noise from a realistic 21-cm experiment is clearly shown in the middle panel of Figure~\ref{fig:ImageCreation}. All small-scale features below our instrumental resolution are lost, leaving only the moderate to large-scale information. Importantly though, in terms of reionisation studies, for the most part the ionisation structure (ionised regions) are still identifiable. This implies that we would expect low $j$ scattering coefficients to be heavily affected and/or dominated by the instrumental effects, with the higher $j$ scattering coefficients being more sensitive to the 21-cm signal. However, this assumes perfect foreground removal (i.e. no loss of structural information). In the final panel, we see the impact of removing these contaminated foreground `wedge' modes which notably further reduces the amplitude and structural information in the resultant mock image. 

\subsection{Evolution of WST scattering coefficients under instrumental effects}

\begin{figure*} 
	\begin{center}
		\includegraphics[trim = 1cm 0.6cm 0cm 0.5cm, scale = 0.7]{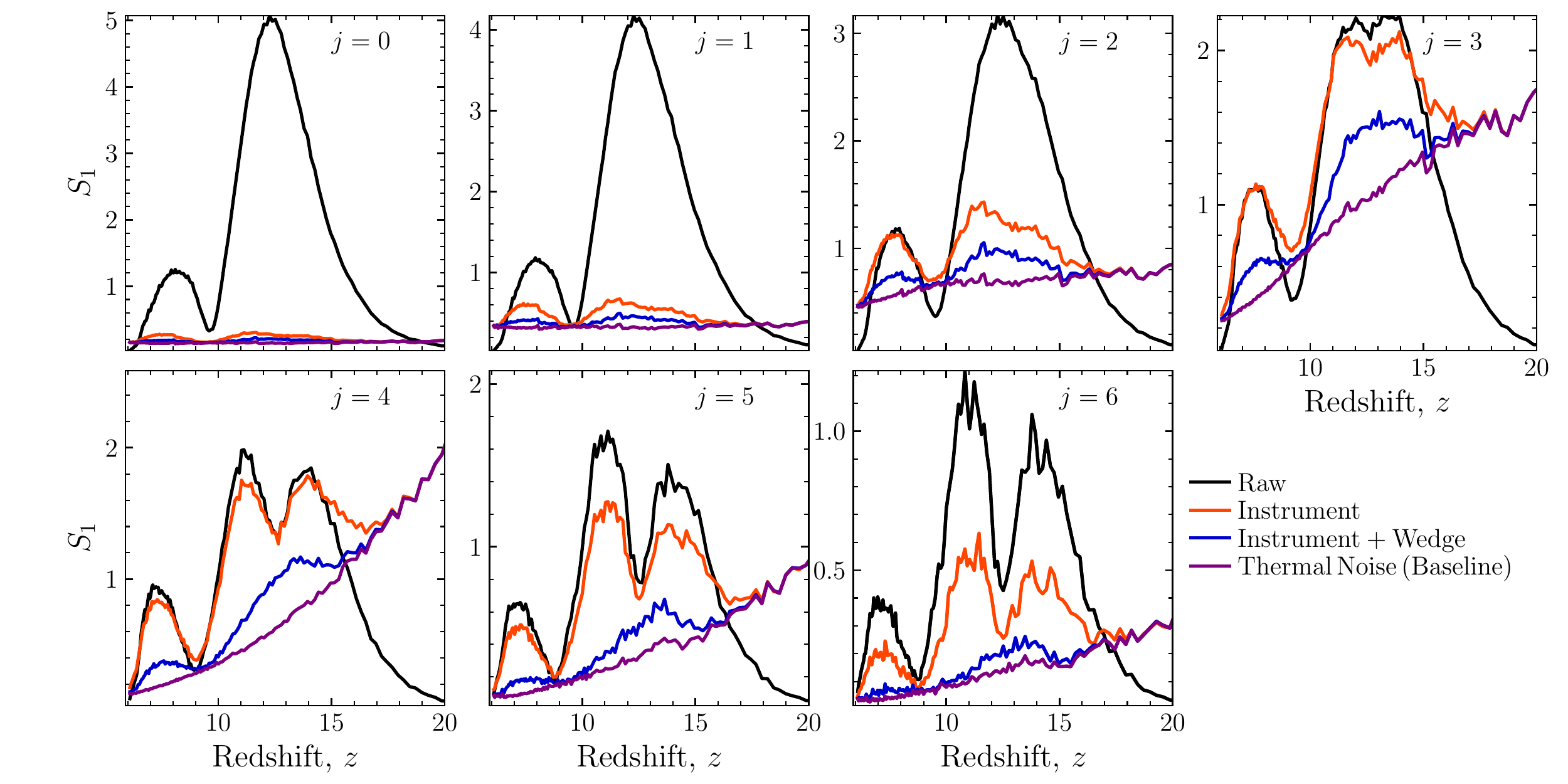}
	\end{center}
\caption[]{The evolution in the $S_{1}$ scattering coefficients in the presence of instrumental effects as a function of redshift for our fiducial model averaged over 30 different simulated 21-cm realisations. We assume a 1000hr observation with SKA1--low. The black curve corresponds to raw, simulated 21-cm images (no instrumental effects), the red curve includes thermal noise and the SKA $uv$-coverage, the blue curve includes thermal noise, the SKA $uv$-coverage and the removal of foreground contaminated (`wedge') modes and finally the purple curves are pure thermal noise realisations only (including SKA1--low resolution).}
\label{fig:S1_Instrument}
\end{figure*}

Previously, in Section~\ref{sec:S-evolution} we explored the evolution of the WST scattering coefficients as a function of redshift to ascertain the utility of the WST applied to the 21-cm signal. Here, we return to this, now investigating the evolution of the scattering coefficients from realistic mock 21-cm images from an experiment such as the SKA1--low. Unlike previously, where we explored four different astrophysical models, here we only focus on our fiducial model, as we are specifically interested in how the instrumental effects impact our ability to infer information on reionisation. Further, we apply the WST to 30 different realisations of our 21-cm signal from our fiducial model to obtain the mean evolution in the scattering coefficients. This choice is motivated by: (i) taking into account the sample variance, which will be necessary for our astrophysical forecasting in the next section and (ii) to better estimate the mean thermal noise obtained from various thermal noise realisations used in our mock images.

In Figure~\ref{fig:S1_Instrument}, we compare the mean evolution of all $S_{1}$ scattering coefficients after applying the WST to the 21-cm signal. The black curves correspond to the WST applied directly to the raw, simulated 21-cm signal (i.e. no instrumental effects). The red and blue curves include both thermal noise and the expected SKA1--low $uv$-coverage, with the latter also including the removal of foreground contaminated `wedge' modes. Finally, to understand the noise floor from thermal noise, the purple curve corresponds to the mean evolution in the $S_{1}$ scattering coefficients including only thermal noise (i.e. no 21-cm signal). This effectively sets the lower bound for the amplitude of the scattering coefficients, with the cosmological information needing to be above this level to be differentiable from the noise.

As established previously, by including thermal noise and the expected SKA1--low $uv$-coverage, the amplitude of the $S_{1}$ coefficients for the low $j$ scattering coefficients are considerably impacted. We lose most of the signal on spatial scales below the SKA resolution (as seen by the precipitous drop in amplitude for $j\leq2$). As we increase the spatial scales of our wavelet filters (increase $j$), the impact of the instrumental effects are minimised. For $j\gtrsim3$, the scattering coefficients including the instrumental effects closely match those from the raw simulation output, and are still sensitive to the three peaks denoting the various epochs of the cosmic 21-cm evolution (EoR, EoH and WF-coupling). Thus, the WST is relatively unaffected by instrumental effects on moderate to large spatial scales (as is the case for the 21-cm PS, see e.g. \citealt{Greig:2015p3675}) for realistic images of the 21-cm signal. It is these scales that are important for astrophysical parameter inference.

\begin{figure*} 
	\begin{center}
		\includegraphics[trim = 1cm 0.8cm 0cm 0.5cm, scale = 0.48]{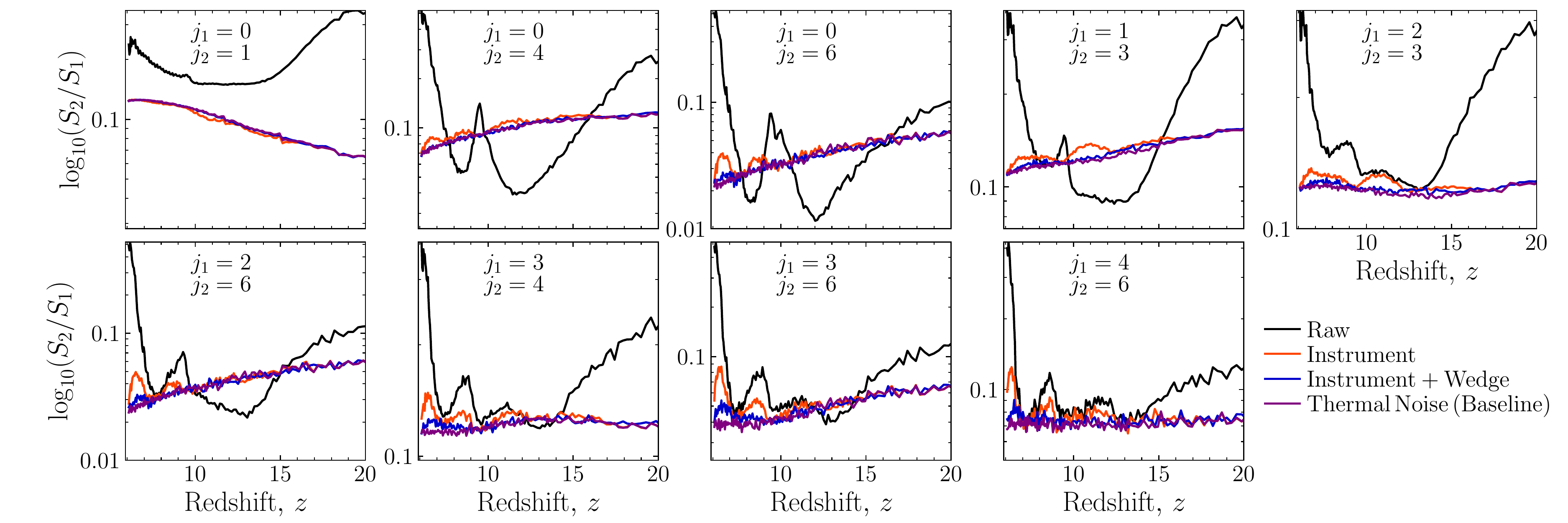}
	\end{center}
\caption[]{The same as Figure~\ref{fig:S1_Instrument}, except for the all the de-correlated $S_{2}$ coefficients. Here, we only present a sub-set of all possible scattering coefficients.}
\label{fig:S2_Instrument}
\end{figure*}

Once we remove the foreground `wedge' modes, the amplitude of the $S_{1}$ scattering coefficients is drastically reduced across all spatial ($j$) scales. Further, we can no longer differentiate between the two peaks in the cosmic 21-cm signal arising from the EoH and WF-coupling eras. Nevertheless, despite the amplitude of the $S_{1}$ coefficients dropping by $\sim 2- 3$ relative to the red curves, the amplitude remains above that set by the thermal noise floor. Though, it does reach the noise floor earlier and for a wider range of redshifts. Importantly, this implies that cosmological information is still accessible from the $S_{1}$ scattering coefficients applied to realistic mock images of the 21-cm signal.

In Figure~\ref{fig:S2_Instrument}, we perform the same comparison, this time for the de-correlated $S_{2}$ coefficients. Here, we specifically limit ourselves to a small subset of $S_{2}$ coefficients noting that many exhibit similar behaviours as seen in Figure~\ref{fig:S2Coefficients}. Similar to above, we find that for any $S_{2}$ coefficients on small spatial scales (i.e. low $j_{1}$ or $j_{2}$) the instrumental effects dominate over the cosmological signal. Similarly, on large spatial scales, we find that the $S_{2}$ coefficients, even in the presence of all sources of instrumental effects, remain above the thermal noise level implying the possibility of being able to  detect the non-Gaussian information from the 21-cm signal in amongst the noise. However, this appears to only be the case at $z<10$, and predominately at $z\lesssim8$. This implies that most of the additional non-Gaussian information that we can obtain from realistic 21-cm images using the WST will come during the EoR, where the non-Gaussianities are strongest due to the complex ionisation morphology. Finally, we re-emphasise that here we are presenting the de-correlated $S_{2}$ coefficients to highlight the non-Gaussian information beyond that held by the $S_{1}$ coefficients. As discussed previously, the actual $S_{2}$ coefficients are larger in amplitude as they retain the spatial information held by the $S_{1}$ coefficients.

\section{Astrophysical parameter forecasts} \label{sec:forecasts}

Thus far, we have only qualitatively explored the WST applied to the 21-cm signal, gaining physical insight and intuition by comparing the evolution of the scattering coefficients as a function of redshift. In this section, we perform a quantitative analysis of the WST in the context of astrophysical parameter inference, determining how well we can constrain our input astrophysical model using our realistic mock images of the 21-cm signal from the SKA1--low. For the remainder of this work, we focus on the fiducial model outlined in Section~\ref{sec:setup}.

While it is plausible to explore the WST applied directly to the 21-cm signal from 3D simulations in a fully Bayesian framework using \cmmc{} \citep{Greig:2015p3675,Greig:2017p8496,Greig:2018,Park:2019}, for simplicity we instead perform a Fisher Matrix analysis. Primarily, this choice is motivated by several uncertainties in the construction of the likelihood function for the WST, which can be more trivially accounted for with a Fisher Matrix. For example, determining the cosmic (sample) variance (and associated covariance matrix) for each of the scattering coefficients requires numerical estimation from many realisations. While this is tractable for the Fisher Matrix, as we only need to estimate it for our mock observation, for direct MCMC we would need to determine this for each set of sampled astrophysical parameters. In the future we will return to this and perform a more robust exploration of the WST with \cmmc{}.

\subsection{Fisher Matrices}

The Fisher Information Matrix \citep{Fisher:1935} estimates the amount of information contained about a model parameter, $\theta_{i}$, given some observational measurement. In our case, we can write the Fisher Matrix as,
\begin{eqnarray} \label{eq:Fisher}
\boldsymbol{\mathsf{F}}_{ij} \equiv - \left\langle  \frac{\partial^{2}{\rm ln}(\mathcal{L})}{\partial \theta_{i} \partial\theta_{j}} \right\rangle = \sum_{z} \frac{\partial^{2}}{\partial\theta_{i}\partial\theta_{j} }\boldsymbol{\mathsf{S}}^{\boldsymbol{\mathsf{T}}}\cdot\boldsymbol{\mathsf{\Sigma}}^{-1}\cdot \boldsymbol{\mathsf{S}},
\end{eqnarray} 
where we sum over the redshifts, $z$, of each individual 2D image, $\mathcal{L}$ is the likelihood function (the probability density of the model given our input parameters), $\boldsymbol{\mathsf{\Sigma}}$ is the covariance matrix containing the cosmic variance and thermal noise errors on each of the scattering transform coefficients and $\boldsymbol{\mathsf{S}}$ contains all possible scattering coefficients (i.e. first and second order). In defining the Fisher Matrix in this way, we assume that each observational measurement of the scattering transform coefficients are statistically independent in redshift (frequency). Thus, we sum the information over all possible redshifts for which we extract a 21-cm image.

We obtain parameter forecasts for our individual astrophysical parameters by calculating the covariance matrix, $\boldsymbol{\mathsf{C}}_{ij}$, which is simply given by the inverse of the Fisher Matrix. The resultant confidence intervals obtained through this method are considered as lower bounds on the true precision (i.e. best possible constraints) given the observational setup.

\subsection{Comparing the WST to the 21-cm Power Spectrum}

First, in order to be able to provide a baseline from which to compare the performance of the WST against, we compute the Fisher Matrix using the 3D spherically averaged 21-cm PS. While we could theoretically compare the 21-cm PS extracted from a 2D 21-cm image to more directly compare to the 2D WST, we refrain from doing so as it is not what will be explored in practise. Instead, in this work, we want to compare the WST to the established literature, for which the 3D spherically averaged 21-cm PS is the benchmark as it is the most well explored and understood. Alternatively, we could instead apply the WST in 3D, however, interpreting the scattering coefficients will be less trivial owing to the redshift dependence along the line-of-sight. However, in future, we will return to this. 

It is important to note that in making this choice we cannot perform a straight-forward like-for-like comparison. As we will be comparing the 2D WST to the 3D spherically averaged PS, inherently there will be more statistical information available with the 3D PS as it will average over more structural information also leading to a smaller sample variance error due to the extra information. Further, by measuring the 21-cm signal across a bandwidth with the 3D PS, the instrumental errors (i.e. noise) will also reduce down for the PS compared to that of a single 2D image. Nevertheless, the comparison serves to highlight our understanding that astrophysical inference from the first-order scattering coefficients behaves in a similar way to the 21-cm PS. Further, from this, we can then determine how well things improve when considering the additional non-Gaussian information from the $S_2$ coefficients in Section~\ref{sec:fullWST}.

\subsubsection{21-cm Power Spectrum Fisher Matrix}

For the mock observation using the 21-cm PS, we closely follow the approach of \citet{Park:2019}. We break the observed 21-cm light-cone along the line-of-sight (redshift) direction into equal comoving volumes equivalent in size to our 250 Mpc simulation. For each chunk of the 21-cm light-cone we then compute the 3D spherically averaged 21-cm PS resulting in 12 independent measurements of the 21-cm PS\footnote{The central redshift for each of these individual pieces of the 21-cm light-cone correspond to: $z=6.3, 7.0$, 7.7, 8.6, 9.6, 10.8, 12.2, 13.8, 15.8, 18.1, 21.0 and 24.7.} between $z\sim5.9 - 27.4$. Using \textsc{\small 21cmSense}, we determine the observational errors (sample variance and thermal noise) on our mock observation of the 21-cm PS assuming a 1000 hr observation with the SKA1--low. Combining these sources of error in quadrature, we then compute our Fisher Matrix for the 21-cm PS using a similar format to that expressed in Equation~\ref{eq:Fisher}. Following \citet{Park:2019}, we restrict our Fisher Matrix computation from the 21-cm PS to eight linear\footnote{This choice to linearly sample the bins in Fourier space is to increase the sampling towards moderate to large spatial scales (i.e. small $k$) where there is increased sensitivity to the astrophysical information. However, over the range in which we compute this, the differences between this and logarithmically sampling are minimal.} spaced Fourier bins between $k=0.1 - 1.0$~Mpc$^{-1}$, with the lower limit set by astrophysical foreground contamination and the upper limit being shot-noise within the 21-cm simulations.

Importantly for this comparison, unlike \citet{Park:2019} we do not consider the modelling uncertainty term. This approximately quantifies how much the simulated 21-cm PS obtained from a semi-numerical simulation differs to that from a full radiative transfer simulation using the same initial conditions. In order to be able to perform a fair comparison between the WST and the 21-cm PS we must ignore this source of error as we do not have an equivalent estimate for the modelling uncertainty for the WST. In the future, we will return to this. Additionally, this implies we cannot perform a like-for-like comparison between our Fisher Matrix estimates for the 21-cm PS and the direct MCMC performed in \citet{Park:2019} (their 21-cm only case). However, while not shown here, we verified that when including this modelling uncertainty, our Fisher Matrix framework recovered comparable constraints to those presented in \citet{Park:2019}. Equally, we recovered similarly good agreement with the 21-cm Fisher Matrix code, 21cmFish (Mason et al., in prep).

\subsubsection{$S_{1}$ scattering coefficients Fisher Matrix}

In Section~\ref{sec:s1}, we highlighted that the first-order scattering coefficients are analogous to the 21-cm PS. Here, we quantify this by estimating a Fisher Matrix using just the $S_{1}$ coefficients. In order to be able to perform this comparison, we limit the number of 21-cm images used for the WST to ensure we have a comparable number of independent measurements. Specifically, we extract the scattering coefficients from 21-cm images extracted at the central redshift (frequency) of the 12 equal co-moving cubes of the 21-cm light-cone see previous section. Further, for our simulation setup, we have seven $S_{1}$ coefficients, compared to the eight 21-cm PS bins per redshift. Thus, we have a comparable number of data-points across the two approaches.

For this comparison, we only consider the impact of the SKA1--low instrument resolution and thermal noise on our mock 21-cm images (i.e. following Section~\ref{sec:instrument}) leaving an exploration of the impact of excising the foreground contaminated `wedge' modes until Section~\ref{sec:Fisher-wedge}. To compute the Fisher Matrix for the $S_{1}$ coefficients (and subsequently all WST Fisher Matrices), we first determine the mean scattering coefficients averaged across our 30 independent realisations of the 21-cm signal. We then take the numerical derivatives of these mean coefficients with respect to our astrophysical parameters. This is chosen to minimise any potential numerical noise artefacts in the scattering coefficients owing to our chosen image size (i.e. 250 Mpc). In averaging across these 30 realisation, we additionally determine our error covariance (sample variance) for the scattering coefficients. We then add this in quadrature with the thermal noise error for the scattering coefficients obtained from 21-cm images to obtain an estimate of the total noise, $\sigma$ from Equation~\ref{eq:Fisher}. This thermal noise error is determined by averaging different noise realisations (i.e. no 21-cm signal) and applying the same SKA1--low $uv$-coverage, amounting to the mean uncertainty on the scattering coefficients from thermal noise only (no 21-cm signal).

\begin{figure*} 
	\begin{center}
		\includegraphics[trim = 0.2cm 0.6cm 0cm 0.4cm, scale = 1.1]{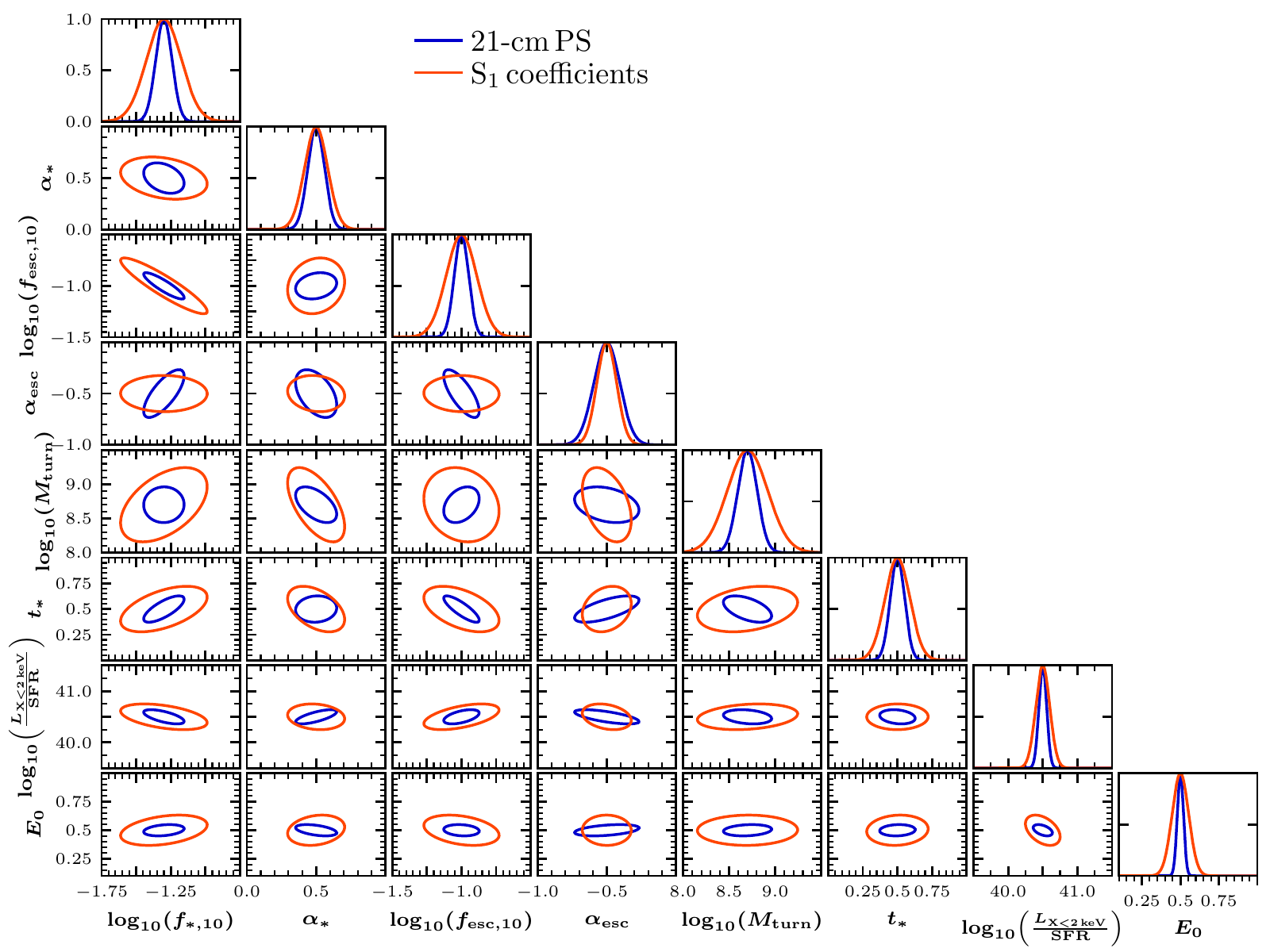}
	\end{center}
\caption[]{The recovered confidence intervals for our astrophysical parameters from the fiducial model estimated using Fisher Matrices assuming a 1000 hr observation with SKA1--low. The one dimensional probability distribution functions for each parameter are shown along the diagonal, whereas the two dimensional confidence intervals are shown below. Here, these contours correspond to the 95 percentiles. The blue curves correspond to the constraints using the 21-cm power spectrum while the red curves correspond to the $S_{1}$ scattering coefficients obtained from the 2D WST.}
\label{fig:Fisher1}
\end{figure*}

\subsubsection{Comparing astrophysical parameter constraints}

\begin{table*}
\begin{tabular}{@{}lccccccccc}
\hline
  & ${\rm log_{10}}(f_{\ast,10})$ & $\alpha_{\ast}$ & ${\rm log_{10}}(f_{\rm esc,10})$ & $\alpha_{\rm esc}$ & ${\rm log_{10}}(M_{\rm turn})$ & $t_{\ast}$ & ${\rm log_{10}}\left(\frac{L_{{\rm X}<2{\rm keV}}}{\rm SFR}\right)$ & $E_0$   \\
               &  &  &  & & $[{\rm M_{\sun}}]$ & &  $[{\rm erg\,s^{-1}\,M_{\sun}^{-1}\,yr}]$ &  $[{\rm keV}]$ \\
\hline
\vspace{0.8mm}
Mock Obs. & $-1.30$ & $0.50$ & $-1.00$ & $-0.50$ & $8.7$ & $0.5$ & $40.50$ &  $0.50$\\
\hline
\vspace{0.8mm}
21-cm PS & $\pm0.06$ & $\pm0.06$ & $\pm0.05$  & $\pm0.09$  & $\pm0.10$  &  $\pm0.05$   &  $\pm0.06$   &  $\pm0.02$ \\
\hline
\vspace{0.8mm}
Instrumental effects &  &  &   &   &   &     &     &   \\
$S_{1}$ & $\pm0.13$ & $\pm0.08$ & $\pm0.11$  & $\pm0.07$  & $\pm0.22$  &  $\pm0.09$   &  $\pm0.10$   &  $\pm0.05$ \\
\vspace{0.8mm}
$S_{2}$ (de-corr.) & $\pm0.18$ & $\pm0.12$ & $\pm0.15$  & $\pm0.14$  & $\pm0.69$  &  $\pm0.15$   &  $\pm0.30$   &  $\pm0.13$ \\
\vspace{0.8mm}
$S_{1}+S_{2}$ & $\pm0.07$ & $\pm0.05$ & $\pm0.07$  & $\pm0.05$  & $\pm0.13$  &  $\pm0.05$   &  $\pm0.06$   &  $\pm0.04$ \\
\hline
\vspace{0.8mm}
Wedge removed &  &  &   &   &   &     &     &   \\
\vspace{0.8mm}
$S_{1}$ & $\pm0.20$ & $\pm0.13$ & $\pm0.22$  & $\pm0.12$  & $\pm0.54$  &  $\pm0.11$   &  $\pm0.24$   &  $\pm0.16$ \\
\vspace{0.8mm}
$S_{1}+S_{2}$ & $\pm0.12$ & $\pm0.07$ & $\pm0.13$  & $\pm0.07$  & $\pm0.31$  &  $\pm0.07$   &  $\pm0.14$   &  $\pm0.07$ \\
\hline
\vspace{0.8mm}
Wedge removed &  &  &   &   &   &     &     &   \\
\vspace{0.8mm}
(different cadence) &  &  &   &   &   &     &     &   \\
\vspace{0.8mm}
$S_{1}+S_{2}$ (2 MHz) & $\pm0.05$ & $\pm0.03$ & $\pm0.04$  & $\pm0.03$  & $\pm0.13$  &  $\pm0.03$   &  $\pm0.06$   &  $\pm0.03$ \\
\vspace{0.8mm}
$S_{1}+S_{2}$ (5 MHz) & $\pm0.07$ & $\pm0.04$ & $\pm0.07$  & $\pm0.04$  & $\pm0.20$  &  $\pm0.04$   &  $\pm0.09$   &  $\pm0.05$ \\
\vspace{0.8mm}
$S_{1}+S_{2}$ (10 MHz)  & $\pm0.10$ & $\pm0.05$ & $\pm0.09$  & $\pm0.06$  & $\pm0.26$  &  $\pm0.06$   &  $\pm0.12$   & $\pm0.06$ \\
\hline

\end{tabular}
\caption{The recovered $1\sigma$ confidence intervals for our astrophysical parameters from the 21-cm power spectrum, various combinations of the scattering transform coefficients ($S_{1}$, $S_{2}$ and $S_{1}+S_{2}$) and different observational effects. In all, we assume a 1000 hr observation with the SKA1-low including its finite $uv$-coverage (instrumental resolution) and thermal noise. We additionally consider the case where we also remove contaminated foreground `wedge' modes before analysing our 21-cm images (wedge removed). Finally, we consider altering the frequency cadence for the sampling of the 21-cm images used in our mock observation (2, 5 and 10~MHz). See text for further details.}
\label{tab:Results1}
\end{table*} 

In Figure~\ref{fig:Fisher1} we compare the resultant constraints on our astrophysical parameters from our fiducial model from the 21-cm PS (blue) and the $S_{1}$ scattering coefficients (red). Along the diagonal are the marginalised one dimensional probability distribution functions (PDFs), while the remaining panels are the recovered two dimensional PDFs (95 percentiles). We summarise the resultant $1\sigma$ confidence intervals on our individual astrophysical parameters in the top two rows of Table~\ref{tab:Results1}. 

Overall, we find that the 21-cm PS recovers our astrophysical parameters at $\sim1.5-2$ times higher precision than the 2D WST. However, this was to be expected given that we are comparing the 3D spherically averaged 21-cm PS to the 2D WST applied to 21-cm images. Thus, while in this scenario we have a similar number of independent data-points (redshifts and filtering scales), the 3D PS contains additional redshift evolution information by averaging information along the line-of-sight direction, which is crucial for constraining our astrophysical parameters. Further, the 2D images of the 21-cm signal used in the WST analysis contain larger sources of error, namely larger sample variance (as we are only averaging over a 2D image rather than a 3D cube for the 21-cm PS) and higher thermal noise as we do not average across the bandwidth (line-of-sight direction). Theoretically, for this comparison, we could reduce either of these sources of noise, which would notably improve the performance of the WST relative to the 21-cm PS, however, the actual performance is less relevant, instead it is about gaining intuition about the WST. Taking into the account caveats, quite clearly this demonstrates that the information contained in the $S_{1}$ coefficients is equivalent to the 21-cm PS.

Importantly, from Figure~\ref{fig:Fisher1} we mostly recover similar astrophysical degeneracies for the two statistical methods. This confirms our intuition that the $S_{1}$ scattering coefficients behave in a similar way to the 21-cm PS, and follows from \citet{Cheng:2020}, who noted the similarities between the cosmological constraints obtained from weak lensing from the PS and the $S_{1}$ coefficients. However, we do observe some notable changes in the orientations for some ellipses potentially implying different sensitivities to the underlying astrophysics. These likely arise from the fact that the 21-cm PS is the square of the modulus of the convolved field whereas the $S_{1}$ coefficients are simply the modulus, allowing some differences to occur. Note, in principle it is also possible to combine both statistics to further improve the constraining power, however, in this work we refrain from doing so focussing more on the comparison between the two approaches.

\subsection{Full WST} \label{sec:fullWST}

\begin{figure*} 
	\begin{center}
		\includegraphics[trim = 0.2cm 0.5cm 0cm 0.4cm, scale = 1.1]{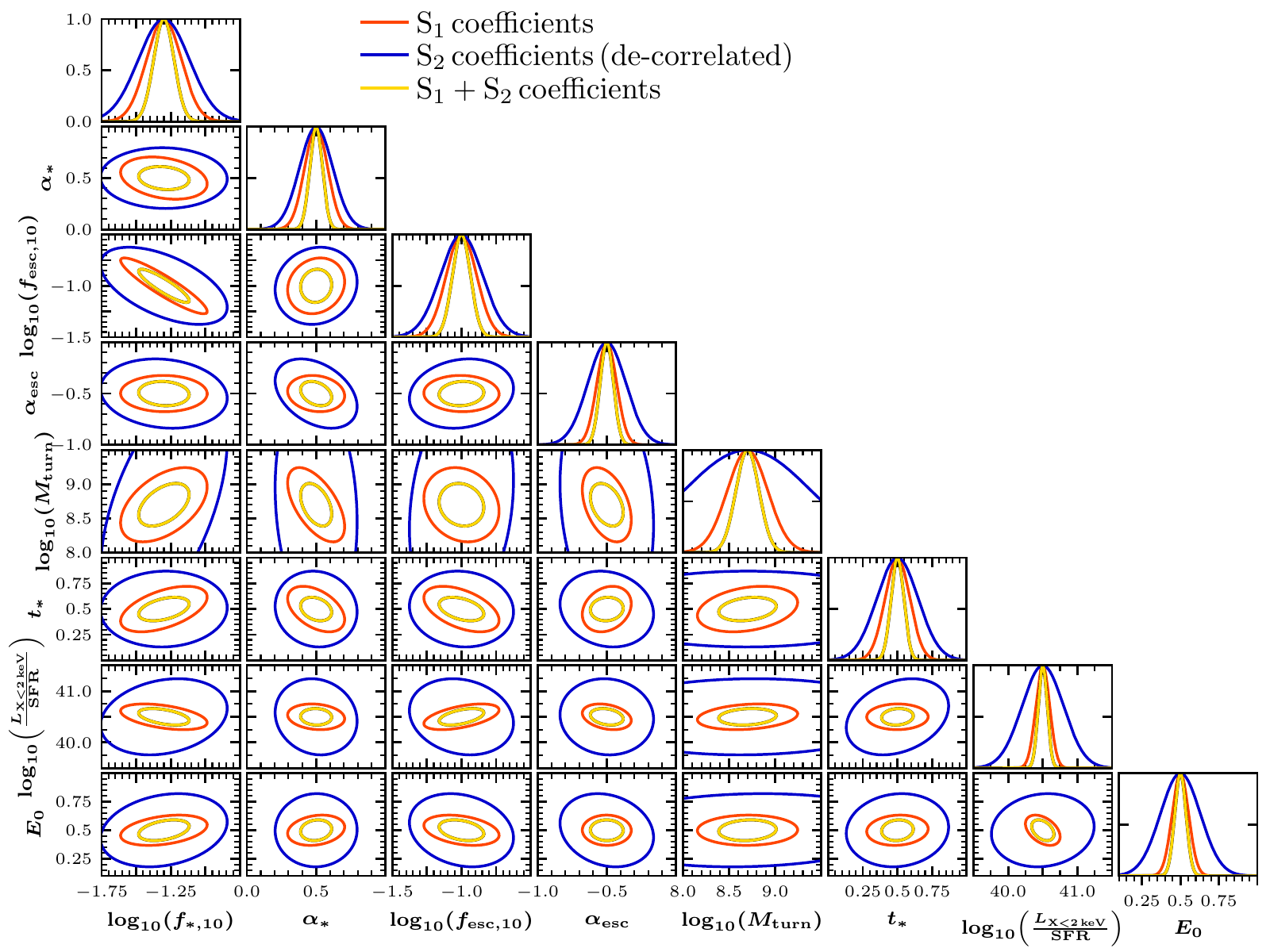}
	\end{center}
\caption[]{Similar to Figure~\ref{fig:Fisher1}, except we compare the astrophysical parameter constraints obtained from just the $S_{1}$ coefficients (red), the de-correlated $S_{2}$ coefficients (blue) and the full combination of $S_{1}+S_{2}$ coefficients (yellow).}
\label{fig:Fisher2}
\end{figure*}

The true value of the WST for astrophysical parameter inference comes from the inclusion of the additional non-Gaussian information obtained in the second-order coefficients. Therefore using the same setup as in the previous section we now highlight the available improvement when including this non-Gaussian information. In Figure~\ref{fig:Fisher2} we compare the one and two dimensional marginalised PDFs for just the $S_{1}$ coefficients (red curve), the de-correlated $S_{2}$ coefficients (blue curve) and then combining all coefficients (i.e. $S_{1} + S_{2}$; yellow curve). Additionally, we provide the recovered 68th percentiles for each of the astrophysical parameters in Table~\ref{tab:Results1}.

With the addition of the non-Gaussian information following the inclusion of the $S_{2}$ coefficients, we find $\sim1.5-2\times$ improvements for the astrophysical parameter constraints relative to those from just the $S_{1}$ coefficients. To highlight how the second order scattering coefficients adds constraining power, we focus on the de-correlated $S_{2}$ coefficients, which isolate out the non-Gaussian information by averaging out the dependence of the first-order information. These are highlighted by the the blue curves in Figure~\ref{fig:Fisher2}. The astrophysical constraints from just the de-correlated $S_{2}$ coefficients are comparable, though marginally worse than those from $S_{1}$ for all expect $M_{\rm turn}$, $\frac{L_{{\rm X}<2{\rm keV}}}{\rm SFR}$ and $E_{0}$. Instead for these, the constraints are $\sim2-2.5$ times worse than the $S_{1}$ coefficients. However, this is consistent with the picture from \citet{Watkinson:2021} and \citet{Tiwari:2021}, where for a much simpler three parameter reionisation model the constraints from just the isosceles bispectrum (a single configuration of the bispectrum) alone are larger than from the 21-cm PS\footnote{\citet{Tiwari:2021} also consider the scenario of all possible unique bispectrum configurations, showing an improvement over the 21-cm PS. In doing so, this uses more information than available from the $S_{2}$ coefficients as these are coarsely binned over a few $j$ spatial configurations only. However, computing the likelihood for the bispectrum is considerably more challenging owing to the noisiness of the statistic.}. For $\frac{L_{{\rm X}<2{\rm keV}}}{\rm SFR}$ and $E_{0}$ the reduction in constraining power arises owing to the fact that the 21-cm signal is most sensitive to the X-ray parameters during the Epoch of Heating (i.e.\ $z\gtrsim10$). During this time (see Figure~\ref{fig:S2Coefficients}) the $S_2$ scattering coefficients are considerably less sensitive, thus there is little constraining power on the X-ray parameters from the non-Gaussian information.

Importantly, in several cases the recovered contours from the de-correlated $S_{2}$ coefficients are (or are close to) orthogonal to those from the $S_{1}$ coefficients. This orthogonality indicates that the $S_{2}$ coefficients are sensitive to additional information which is capable of breaking degeneracies amongst the astrophysical parameter constraints from just the $S_{1}$ coefficients alone. This is the value of including the non-Gaussian information from the WST when characterising the 21-cm signal, where we are now additionally measuring the clustering of the sizes of the spatial fluctuations in the 21-cm signal rather than just the sizes of the spatial fluctuations with Gaussian statistics such as the 21-cm PS. Even for parameter combinations where the de-correlated $S_{2}$ coefficients are not orthogonal to the $S_{1}$ constraints, we still see improvements when combining the $S_{1}$ and $S_{2}$ coefficients.

\begin{figure*} 
	\begin{center}
		\includegraphics[trim = 0.2cm 0.5cm 0cm 0.4cm, scale = 1.1]{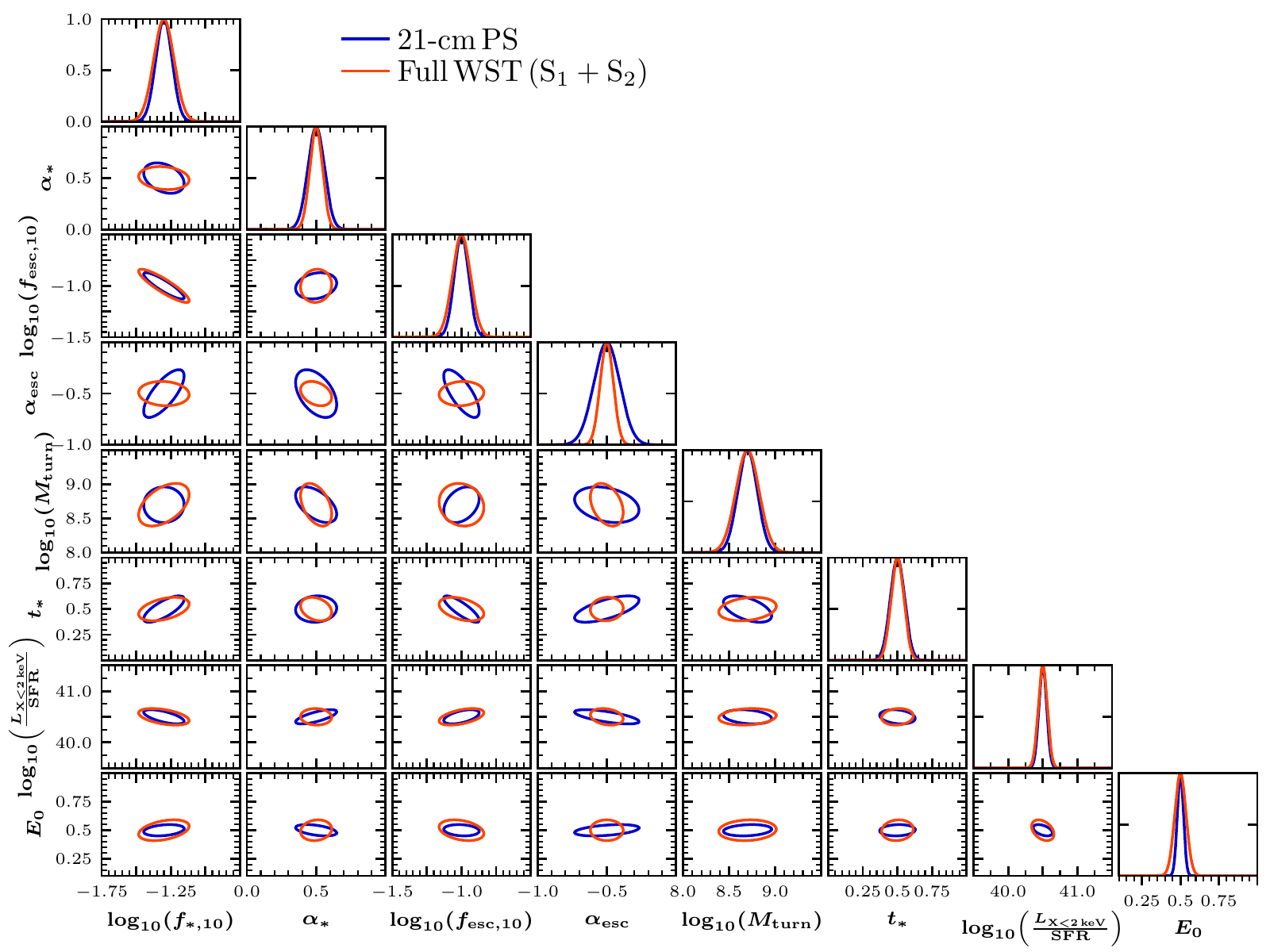}
	\end{center}
\caption[]{Similar to Figure~\ref{fig:Fisher1}, except we compare the astrophysical parameter constraints obtained from full 2D WST (i.e. the $S_{1}+S_{2}$ coefficients; red curves) relative to those obtained from the 3D 21-cm PS (blue curve).}
\label{fig:FisherPS-ST}
\end{figure*}

To fully appreciate the improvement in the constraining power with the 2D WST when using the non-Gaussian information, in Figure~\ref{fig:FisherPS-ST}, we compare the marginalised PDFs for the full 2D WST (red curves) and the 3D 21-cm PS (blue curves). From this we clearly see that the full 2D WST is now comparable to that from the 21-cm PS, even marginally improving on the constraining power for $\alpha_{\ast}$ and $\alpha_{\rm esc}$. This highlights the true value of the WST, that despite the fact we are using considerably less information by restricting ourselves to 2D images we are able to recover comparable constraints to statistics utilising the full 3D information. This demonstrates that the full WST is superior to the 21-cm PS, and is the main result of this work.

In theory we could construct the Fisher Matrix for the bispectrum and compare more directly to the WST. However, in this work we refrain from doing so as the bispectrum is a considerably more complex statistic than the 21-cm PS. While analogies can be trivially drawn between the binning of the PS and the information in the $S_{1}$ coefficients (see Figure~\ref{fig:21cmPS_vs_S1}) the same is not true for the bispectrum. A detailed exploration of all possible closure configurations for the bispectrum will be required to ascertain the analogies to the information contained in $S_{2}$ coefficients. Thus, we defer such a detailed exploration to future work. Nevertheless, this highlights another advantage of the WST. That is, the simplicity to which we are able to access and utilise the non-Gaussian information from the higher-order scattering coefficients relative to other statistical methods.

\subsection{Full WST with foreground contaminated modes removed} \label{sec:Fisher-wedge}

\begin{figure*} 
	\begin{center}
		\includegraphics[trim = 0.2cm 0.5cm 0cm 0.4cm, scale = 1.1]{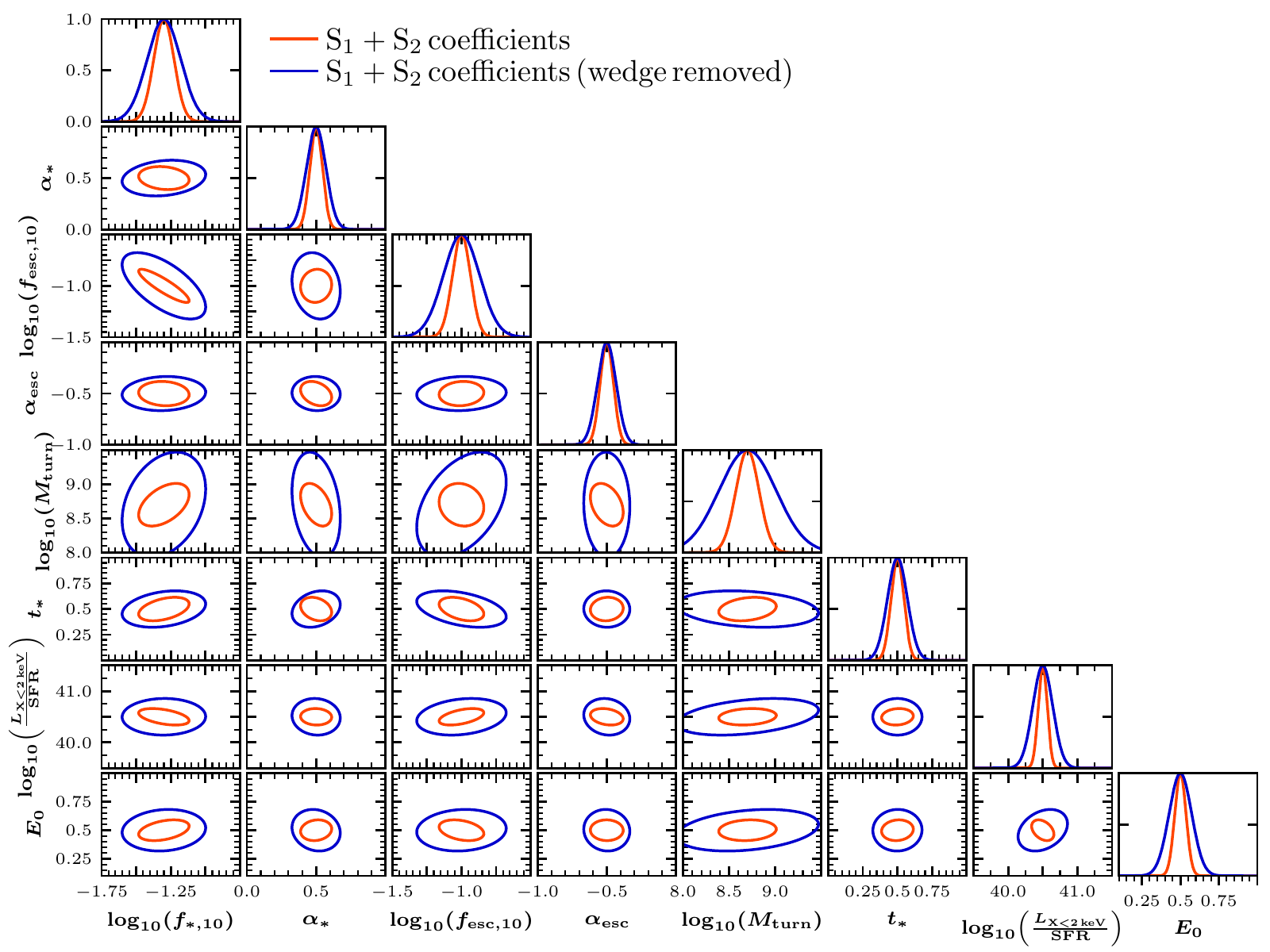}
	\end{center}
\caption[]{Similar to Figure~\ref{fig:Fisher1}, except we explore the impact of removing the contaminated foreground `wedge' modes from our mock 21-cm images using the $S_{1}+S_{2}$ coefficients. The blue curves have these contaminated modes removed, whereas the red curves assumes no foreground contamination (i.e. perfect removal).}
\label{fig:Fisher3}
\end{figure*}

In Figure~\ref{fig:Fisher3} we next compare the one and two dimensional marginalised PDFs from combining the $S_{1} + S_{2}$ coefficients in the presence of contamination by astrophysical foregrounds. The red curves correspond to no contamination (i.e. perfect removal of foreground contamination) whereas the blue curves correspond to the removal of all Fourier modes from within the contaminated foreground `wedge' from our interferometer experiment. We also provide the recovered 68th percentiles for each of the astrophysical parameters in Table~\ref{tab:Results1}.

Following the removal of these contaminated foreground `wedge` modes (see Figure~\ref{fig:ImageCreation} for a visualisation) we find at most a factor of $\sim1.5 - 2$ reduction in the constraining power of our astrophysical parameters.  Nevertheless, this is still a factor of $\sim2$ improvement over that achievable with just the $S_1$ coefficients (i.e. our proxy for the 21-cm PS, see Table~\ref{tab:Results1}). This loss in constraining power following the removal of the wedge modes is to be expected, given the loss of spatial information in the 21-cm signal. While excising these contaminated foreground modes reduces our constraining power from 21-cm images, there are promising approaches in the literature using machine learning to attempt to recover the lost cosmic signal from these foreground contaminated modes \citep[see e.g][]{Gagnon-Hartman:2021}. Thus our ability to constrain astrophysical parameters from images of the 21-cm signal could be further improved by combining both methods. We leave such an investigation to future work. It is also important to note that thus far we have only considered a fairly limited sampling in redshift (frequency) of 2D images of the 21-cm signal. In principle we can mitigate some of these losses in constraining power simply by increasing the number of independent images used in our WST analysis. We explore this in the next section.

\subsection{Observing Cadence}

\begin{figure*} 
	\begin{center}
		\includegraphics[trim = 0.2cm 0.5cm 0cm 0.4cm, scale = 1.1]{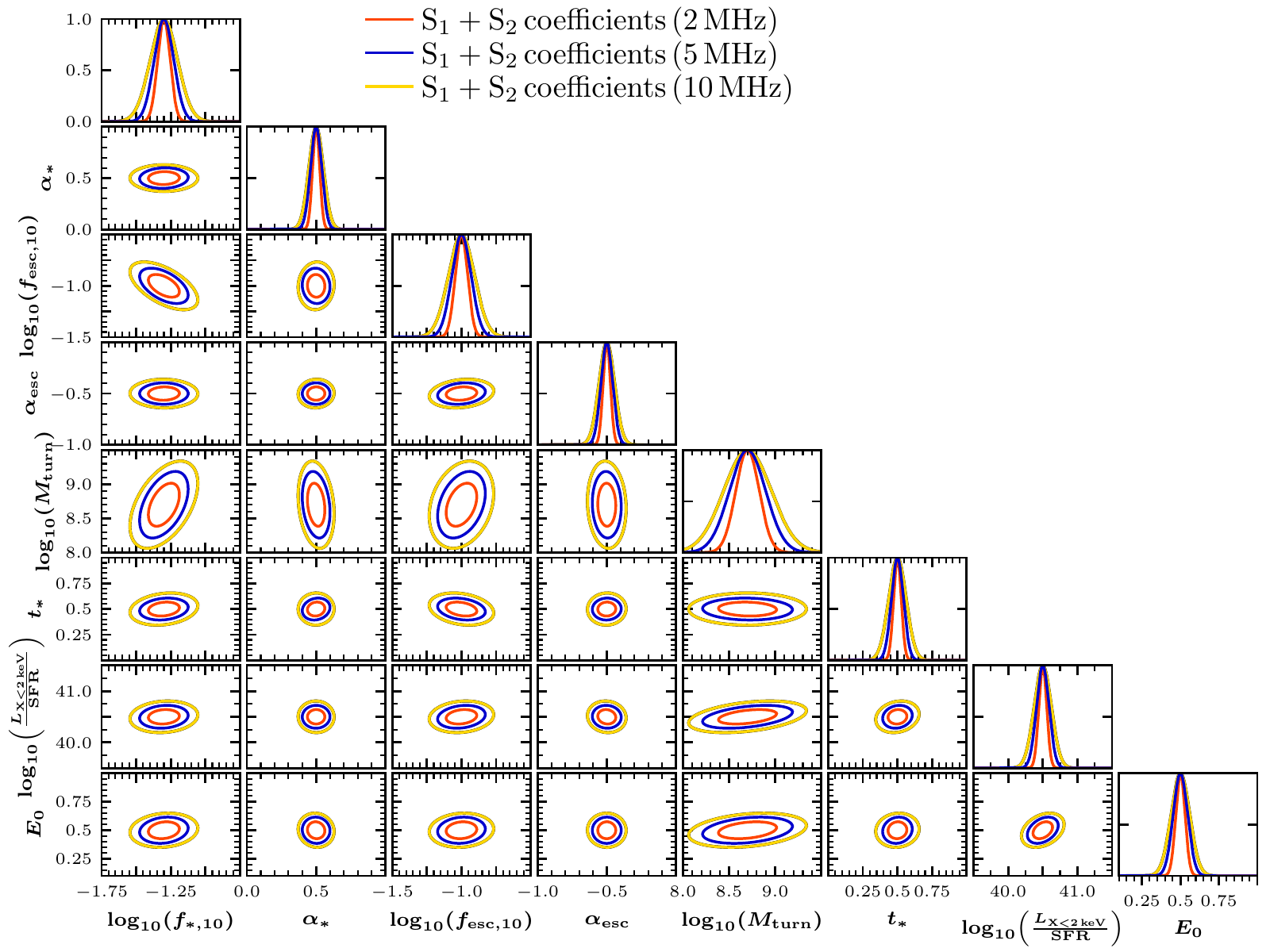}
	\end{center}
\caption[]{Similar to Figure~\ref{fig:Fisher1}, except we compare the astrophysical constraints from combining the $S_{1}+S_{2}$ coefficients for different cadences in frequency between the 21-cm images extracted from the full 21-cm light-cone observed with SKA1--low. We consider cadences of 2 MHz (red), 5 MHz (blue) and 10 MHz (yellow). Here, we consider all instrumental effects including the removal of contaminated foreground `wedge' modes.}
\label{fig:Fisher4}
\end{figure*}

Previously, to aid comparisons between the 21-cm PS and the WST for astrophysical parameter recovery from a mock observation of the 21-cm signal with the SKA1--low we tailored our analysis for the 21-cm PS. That is, we analysed 21-cm images extracted from the central frequency of equal co-moving volumes taken from our simulated 21-cm light-cone (to obtain a similar number of measurements). In effect, this resulted in 12 images which were not equally spaced in frequency, with the largest separation of $\sim18$ MHz during the EoR, down to $\sim10$~MHz at $z\sim20$. In practise, as we observe the full 21-cm light-cone spanning the observing frequency of our interferometer experiment, we can instead extract images at a fixed cadence in frequency. In particular, for our mock observation with the SKA1--low, we consider the 21-cm light-cone spanning $z\sim5.9 - 27.4$, corresponding to a frequency coverage of $\sim206-50$~MHz. Thus, in this section we determine the astrophysical parameter constraints from the WST exploring different cadences in frequency between the 21-cm images extracted from the 21-cm light-cone.

Specifically, we consider three possible cadences in frequency for our 21-cm images: (i) 2 MHz, (ii) 5 MHz and (iii) 10 MHz. That is, we apply the WST to a 2D image of the 21-cm signal extracted from the full light-cone every 2, 5 or 10~MHz. These choices are arbitrary, but serve to demonstrate the possible constraining power available for different cadences. It remains to be seen how many independent images will be achievable with an experiment such as the SKA1--low, but theoretically even higher cadences are plausible. In total, these observational setups result in 79, 32 and 16 independent images of the 21-cm signal extracted from the full observed light-cone, respectively. Note however that beyond about $z\sim15$, the images will be swamped by thermal noise (see Figure~\ref{fig:S1_Instrument} thus not all images add additional constraining power. In all scenarios, we have additional information relative to the previous sections where we only had 12 images corresponding to the central redshift for the 3D 21-cm PS. Further, by now evenly sampling in frequency space, we additionally obtain more measurements during the EoR and EoH unlike previously where we had a cadence of $\sim18$ MHz during these epochs. For all, we consider all observational sources of error with our mock SKA1~--~low observation, instrumental resolution, thermal noise and the removal of foreground contaminated `wedge' modes.

In Figure~\ref{fig:Fisher4} we present the one and two dimensional marginalised PDFs for our astrophysical parameters for these three different cadences in 21-cm images. The red, blue and yellow curves correspond to the 2, 5 and 10~MHz scenarios, respectively. Further, we summarise the 68 percentile constraints on our astrophysical parameters in the final three rows of Table~\ref{tab:Results1}. As expected, following the addition of more independent information by increasing the cadence of the sampled 21-cm images we observe improvements for all astrophysical parameters relative to the cases explored previously. In the case of the 2 MHz sampling ($\sim6.5\times$ more 21-cm images) we recover improvements in our astrophysical parameter constraints by a factor of $\sim2-3$. For 5~MHz sampling ($\sim3\times$ more 21-cm images) we recover improvements of $\sim1.5-2$. Even for the 10~MHz scenario, where the number of images was similar to our previous cases (16 compared to 12) we still note improvements of $\sim10-30$~per cent for most of the astrophysical parameters. These improvements can be attributed to the inclusion of additional 21-cm images (information) during the EoR and EoH relative to previously where the cadence varied. 

Importantly, with cadences of 2 or 5 MHz, the recovered astrophysical parameter constraints are comparable to those from the original 21-cm PS analysis. In fact, for a 2 MHz cadence, the 2D WST outperforms the 3D 21-cm PS for almost all astrophysical parameters by up to a factor of $\sim2$ (for $\alpha_{\ast}$ and $\alpha_{{\rm esc}}$. This, while still only using 2D images which individually contain considerably larger sample variance and thermal noise errors. In principle, we could further reduce the thermal noise in our 2D images by averaging over multiple frequency channels within a fixed bandwidth. Alternatively, we could alter the physical size of our wavelet filters to extract more spatial information. As discussed earlier, we could also instead consider the 3D WST to fully access the available 3D information. Thus, there are many potential avenues to further improve the performance of the WST on astrophysical parameter recovery from the 21-cm signal, some of which we will explore in future. This highlights the unique potential of the WST.

\section{Conclusion} \label{sec:conclusion}

We have introduced the Wavelet Scattering Transform (WST) and applied it to the 21-cm signal for the first time.
The WST \citep{Mallat:2012} convolves a family of wavelet filters of different physical extents and rotations to an input image. The key advantage of the WST comes from its usage of wavelet filters, which preserve locality of the signal both spatially and in frequency. Thus, after successive convolutions and modulus operations applied to the images we are able to extract non-Gaussian information from the 21-cm signal. Importantly, the WST provides a clearly defined series of scattering coefficients, which enables them to be used in a robust statistical analysis.

We performed a qualitative exploration of the WST applied to simulated 21-cm images extracted from the light-cone of four different astrophysical models to gain physical insight. These 21-cm simulations were generated using v3 of \cmfst{} \citep{Mesinger:2007p122,Mesinger:2011p1123,Murray:2020}, specifically the \citet{Park:2019} flexible galaxy parameterisation. These four astrophysical models were chosen to highlight different features in the evolution of the 21-cm signal. We considered a fiducial model, which matches all existing observational constraints, a `cold' reionisation scenario which amplifies the 21-cm signal in the absence of little to no X-ray heating, a model driven by large haloes (resulting in large ionised regions) and an extended model with reionisation driven by faint, low mass haloes (small ionised regions).

Next, we performed a quantitative analysis of the WST by recovering the precision to which we can recover our input astrophysical model parameters from a mock 1000hr SKA1--low observation. Specifically, we used Fisher Matrices and performed several different explorations for which we summarise below:
\begin{itemize}
\item A comparison between the 3D 21-cm PS and the $S_{1}$ scattering coefficients (i.e. Gaussian information) obtained from 2D images of the 21-cm signal. Qualitatively, we find the $S_{1}$ scattering coefficients recover similar astrophysical degeneracies to the 21-cm PS confirming our expectation that the $S_{1}$ scattering coefficients behave similarly to the 21-cm PS. Quantitatively, we found the 21-cm PS outperformed the $S_{1}$ scattering coefficients, however, these differences can be attributed to the fact that we only considered 2D images, where the statistical and instrumental noise is increased relative to the 3D 21-cm PS.
\item The full potential of the WST was highlighted by considering the constraining power when combining the $S_{1}$ and $S_{2}$ scattering coefficients. We found that despite only using 2D information, we could recover comparable constraints to the 21-cm PS, even improving the constraints in some instances. These improvements are comparable to those achievable when combining the 21-cm PS with the bispectrum \cite[see e.g.][]{Watkinson:2021,Tiwari:2021}. The inclusion of the non-Gaussian information provided by the $S_{2}$ coefficients provides additional orthogonal information about the astrophysical parameters breaking degeneracies present when considering only the 21-cm PS. This is the main result of this work, highlighting that that the WST is superior to the 21-cm PS for extracting astrophysical information. Secondly, that it is trivial to measure, interpret and utilise the non-Gaussian information with the WST compared to other statistical measures.
\item We explored the relative loss of constraining power following the removal of foreground contaminated `wedge' modes from our 21-cm images. With the WST ($S_{1} + S_{2}$) we observed at most a factor of $\sim2$ reduction in the constraining power of our astrophysical parameters following the excision of these contaminated modes.
\item A variety of observationally motivated cadences in frequency between successive images extracted from the 21-cm light-cone were considered (arbitrarily chosen to be 2, 5 and 10~MHz). Even with the foreground `wedge' modes removed, we could further notably improve on the recovered precision of our astrophysical parameters by analysing 21-cm images every 2-5~MHz. In doing so, we found we could improve our constraining power on all model parameters by a factor of $\sim2$ compared to our previous scenarios using the 2D WST.
\end{itemize}

While we have demonstrated that the 2D WST has the potential to improve astrophysical parameter recovery from the 21-cm signal relative to the 3D 21-cm PS, thus far we have only explored it in the context of Fisher Matrices, which are known to overestimate the constraining power owing to several simplifying assumptions (e.g. Gaussian errors). These are suitable for an initial exploration, however, in future we will explore the WST in a more robust Bayesian statistical framework using direct MCMC sampling of 21-cm images with 21CMMC to more accurately account for the inherent modelling and observational uncertainties.

Finally, in this work we have focussed on astrophysical parameter inference with the WST. However, there are many other possible applications of the WST to the reionisation epoch. For example, is the WST sensitive enough to detect the non-Gaussian information when the 21-cm signal is below the thermal noise? Doing so could provide the first detection of the 21-cm signal. Alternatively, can the WST be used to reconstruct information that would otherwise be contaminated by astrophysical foregrounds or other instrumental effects. Further, can the WST be used for cross-correlation studies between observed galaxies and the 21-cm signal \citep[e.g.]{Sobacchi:2016,Hutter:2018b,Vrbanec:2020}. We will explore these applications of the WST in future work.

\section*{Acknowledgements}

We thank the anonymous referee for their comments. We also thank Sihao Cheng for early discussions on this work, Steven Murray for discussions related to applying interferometer noise effects and Charlotte Mason for discussions regarding Fisher Matrices. Parts of this research were supported by the Australian Research Council Centre of Excellence for All Sky Astrophysics in 3 Dimensions (ASTRO 3D), through project number CE170100013.  YST acknowledges financial support
from the Australian Research Council through DECRA Fellowship DE220101520. Parts of this work were performed on the OzSTAR national facility at Swinburne University of Technology. OzSTAR is funded by Swinburne University of Technology.

\section*{Data Availability}

The data underlying this article will be shared on reasonable request to the corresponding author.

\bibliography{Papers}

\appendix

\section{Convergence Tests} \label{app:convergence}

\begin{figure*} 
	\begin{center}
		\includegraphics[trim = 1cm 0.6cm 0cm 0.5cm, scale = 0.7]{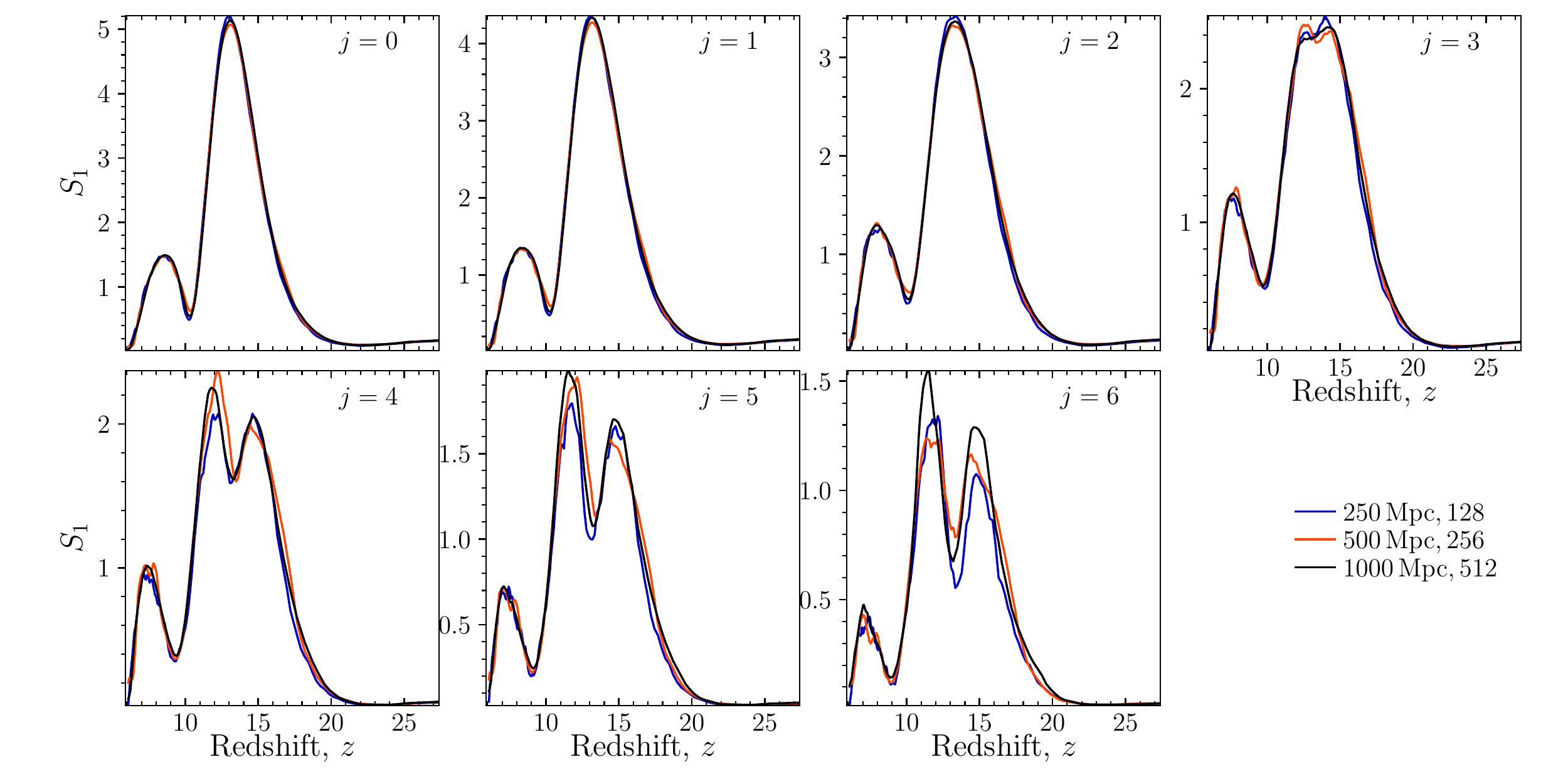}
	\end{center}
\caption[]{The evolution in the $S_{1}$ scattering coefficients for different simulation volumes at fixed resolution. The blue, red and black curves correspond to a transverse side length of 250 (128), 500 (256) and 1000~Mpc (512 voxels).}
\label{fig:Volume}
\end{figure*}

Throughout this work, we consider 21-cm simulations with a transverse comoving length of 250~Mpc and 128 voxels per side length. To ensure our results are consistent across different volumes and simulation resolutions, here we perform some rudimentary convergence tests. Throughout we only show the convergence tests for the $S_{1}$ scattering coefficients, noting we find similar performance for the $S_{2}$ coefficients.

In Figure~\ref{fig:Volume} we explore the redshift evolution of the $S_{1}$ scattering coefficients recovered directly from 21-cm simulations (i.e. no instrumental effects) of varying comoving length for fixed voxel resolution. The blue, red and black curves correspond to transverse lengths of 250 (128), 500 (256) and 1000 Mpc (512 voxels), respectively. Here, we only show $j$'s which are measurable across all three sizes. For all, the recovered evolution in $S_{1}$ scattering coefficients are consistent, irrespective of simulation size (at fixed resolution). The variations across each simulation do increase for larger $j$'s, however, these are consistent to within the sample variance error. That is, for increasing $j$ (i.e. larger physical scales) the smaller simulation sizes will have less information on these scales than the larger volumes, thus the smaller sized simulations are impacted more by sample variance than the larger simulations.

\begin{figure*} 
	\begin{center}
		\includegraphics[trim = 1cm 0.6cm 0cm 0.5cm, scale = 0.7]{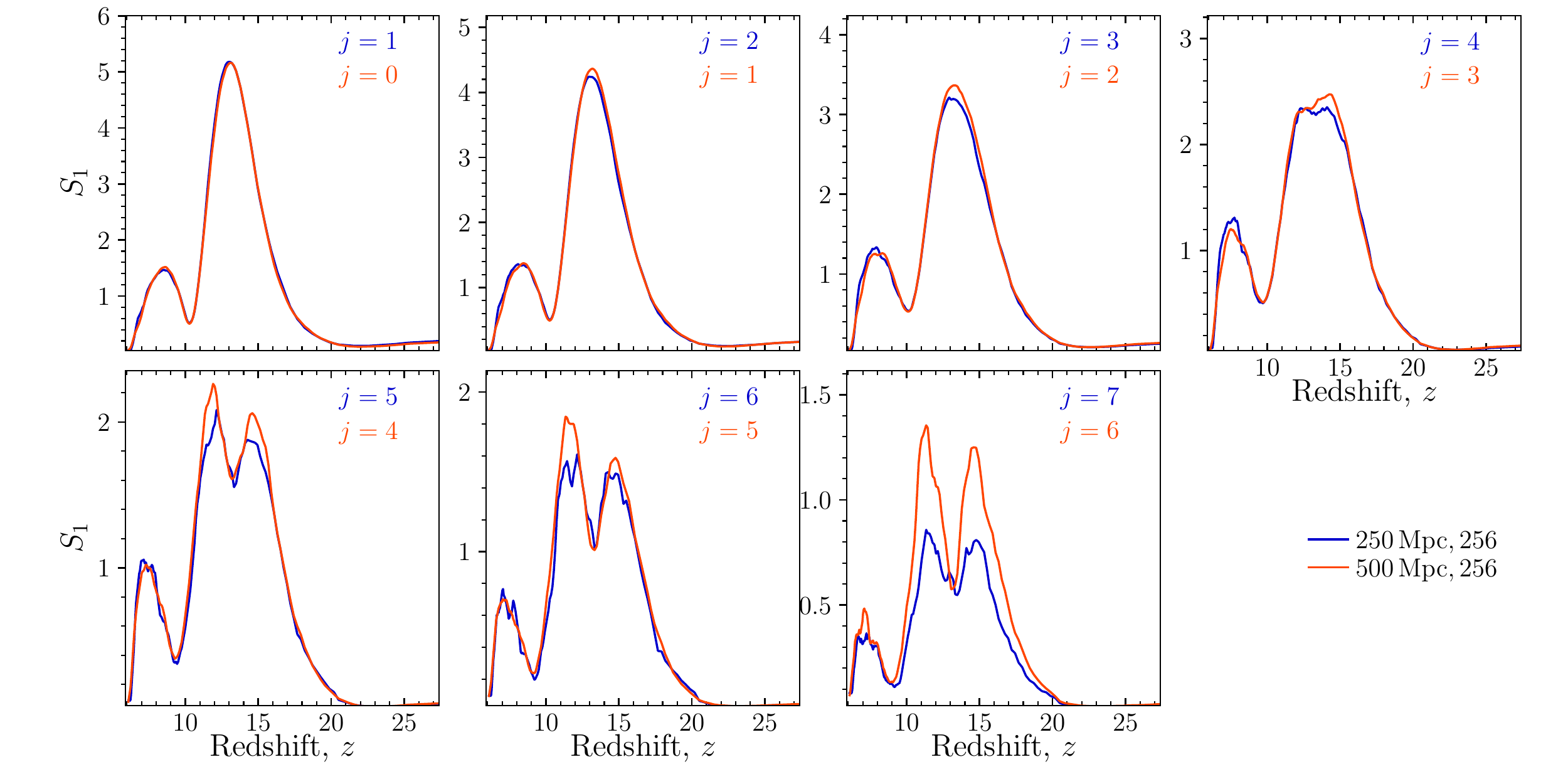}
	\end{center}
\caption[]{The evolution in the $S_{1}$ scattering coefficients for different voxel resolution in our simulations. The blue curves correspond to a simulation of 250 Mpc (256 voxels) whilst the red curve corresponds to 500 Mpc (256 voxels). Note, to accurately compare between resolution we must consider the filters at the same physical extent to ensure the corresponding features are consistent.}
\label{fig:Resolution}
\end{figure*}

In Figure~\ref{fig:Resolution} we instead investigate the redshift evolution of the $S_{1}$ scattering coefficients recovered directly from 21-cm simulations (i.e. no instrumental effects) of varying voxel resolution. Here, the blue and red curves correspond to simulations with 250 (256) and 500 Mpc (256 voxels), corresponding to resolutions of $\sim1$ and $\sim2$~Mpc per voxel. Note, to perform this comparison, the $j$-scales are offset by one to ensure we are comparing to the same physical features in the 21-cm signal (i.e. filters are of the same physical extent). Again, we find the recovered evolution in the $S_{1}$ scattering coefficients are consistent with different voxel resolution. Like previously, we again find the variations increase within increasing $j$-scale, which is due to the increasing sample variance in the higher resolution simulation (i.e. it is at a higher $j$ than the lower resolution simulation).

\section{Smoothing the Scattering coefficients} \label{app:Smoothing}

\begin{figure*} 
	\begin{center}
	 	\includegraphics[trim = 1cm 0.6cm 0cm 0.5cm, scale = 0.7]{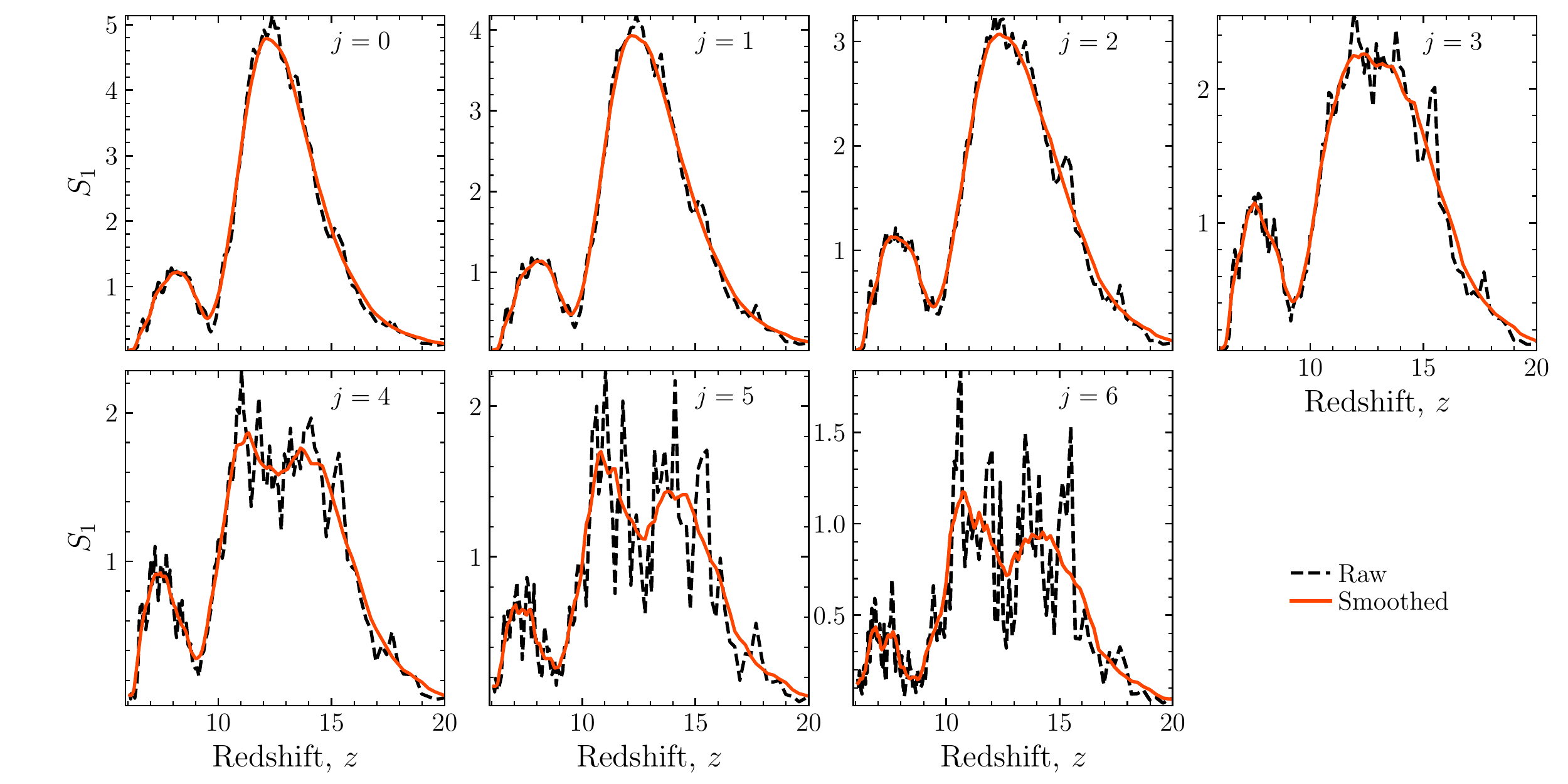}
	\end{center}
\caption[]{Plotting the evolution of the scattering coefficients from the simulated 21-cm images from a single simulation can be relatively noisy (black dashed curve). Thus, we perform a box-car filtering over 10~MHz to smooth out these features (red solid curve). This more clearly highlights the general trend in the scattering coefficient evolution with redshift, and what is performed throughout this work. Here, we present only the $S_{1}$ scattering coefficients, but the same procedure is also applied to the $S_{2}$ coefficients.}
\label{fig:Smoothing}
\end{figure*}

Throughout Section~\ref{sec:WST} we apply the WST to images of the 21-cm signal extracted from a full 3D light-cone from a single simulation to explore the redshift evolution of the scattering coefficients. In particular, we are applying the WST to images that are generated from a single slice in our simulation, corresponding to only $\sim1.5$~comoving Mpc ($\sim100$~kHz). Which is considerably narrower in depth than images that would be obtained from 21-cm interferometer experiments. As such, we experience relatively noisy features in our scattering coefficients when exploring their evolution with redshift, as demonstrated by the black dashed curves in Figure~\ref{fig:Smoothing} for our fiducial astrophysical model. To mitigate these noisy features, we instead perform a box-car filtering in frequency, over a 10 MHz bandwidth. That is, for each scattering coefficient we average over the values of the scattering coefficient $\pm5$~MHz from the current frequency. This smoothing is demonstrated by the red curve in Figure~\ref{fig:Smoothing}, and what is performed throughout this work. In doing so, the general trends for the evolution in the scattering coefficients are more clearly visible.

Importantly, while we could minimise these noisy features by either analysing larger images or increasing the cadence of our light-cone sampling and considering an appropriately narrower box-car filter, the advantage of our current scheme is that it also conveniently mimics the sample variance error. That is, we have only considered one single realisation rather than considering a range of different random initial conditions. For example, in Figure~\ref{fig:S1_Instrument}, we instead show the mean $S_{1}$ scattering coefficients (black solid curve) recovered after averaging across 30 different realisations of our fiducial model, which closely matches the shape and amplitude of the red curve in Figure~\ref{fig:Smoothing}. Thus our current box-car smoothing approach for illustrating the evolution of the scattering coefficients is sufficient for this work.

\section{Randomising phase information} \label{app:Phases}

\begin{figure*} 
	\begin{center}
		\includegraphics[trim = 1cm 0.8cm 0cm 0.5cm, scale = 0.48]{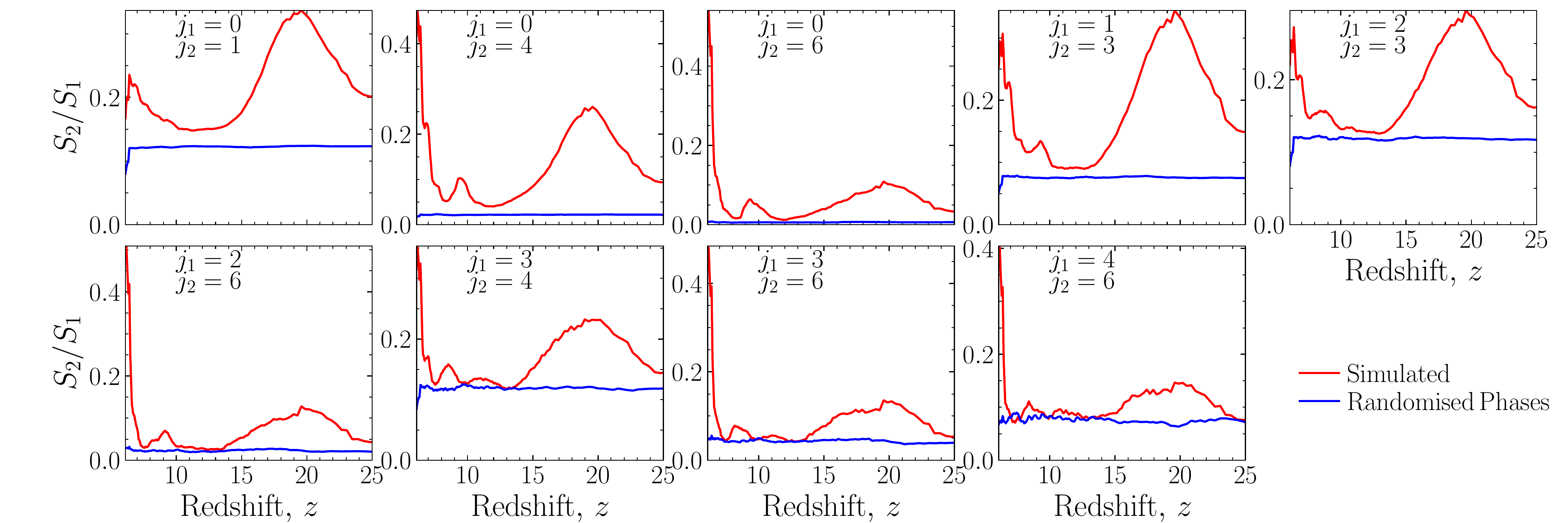}
	\end{center}
\caption[]{A demonstration of the additional information contained in the de-correlated $S_{2}$ coefficients (dividing out the dependence from the $S_{1}$ coefficients) compared to the case when the phase information in the first-order filtered images is randomised. Here, we only show the evolution in a select subset of de-correlated $S_{2}$ coefficients for our fiducial model.}
\label{fig:Phases}
\end{figure*}

In Section~\ref{sec:S2}, we introduced the $S_{2}$ scattering coefficients, highlighting that these contain important non-Gaussian information about the original image. These $S_{2}$ coefficients are correlated with their $S_{1}$ counterparts, thus in Figure~\ref{fig:S2Coefficients} we explored their evolution with redshift with the de-correlated second-order coefficients ($S_{2}/S_{1}$). These de-correlated second-order coefficients showed additional information that we interpret as being the non-Gaussian signal from the input 21-cm image.

As the $S_{2}$ coefficients are obtained by the convolution of the first-order filtered images of the original 21-cm signal, we can verify that the recovered information is truly the non-Gaussian nature of the cosmic signal by randomising the phase information prior to the second convolution by the wavelet filter. In Figure~\ref{fig:Phases} we demonstrate this for all combinations of the de-correlated $S_{2}$ coefficients for our fiducial model. The red solid curves corresponds to the $S_{2}$ coefficients calculated directly from the simulated 21-cm signal, whereas the blue curves correspond to the coefficients after we randomised the phase information in the first-order filtered images before convolution by the second-order family of wavelet filters. Clearly, we see that if we randomise the phase information in the first-order filtered images, the secondary convolution (i.e. $S_{2}$) does not yield any information other than a mean signal. Thus, the $S_{2}$ coefficients are indeed accessing the non-Gaussian information from the input 21-cm signal.

\end{document}